\documentclass[aps,prb,showpacs,twocolumn,superscriptaddress]{revtex4-1}
\usepackage{graphicx,amsmath,amssymb}
\usepackage[usenames]{color}
\usepackage{indentfirst}
\usepackage{float}
\usepackage[T1]{fontenc}
\usepackage[colorlinks=true, citecolor=blue, urlcolor=blue, linkcolor=blue ]{hyperref}
\hypersetup{breaklinks=true}
\begin{document}

\bibliographystyle{apsrev4-1}

\title{Altermagnetism in Heavy Fermion Systems: Mean-Field study on Kondo Lattice}

\author{Miaomiao Zhao}
\affiliation{Key Laboratory of Quantum Theory and Applications of MoE $\&$ School of Physical Science and Technology, Lanzhou University, Lanzhou 730000, People Republic of China}
    \affiliation{Lanzhou Center for Theoretical Physics, Key Laboratory of Theoretical Physics of Gansu Province, Lanzhou University, Lanzhou 730000, People Republic of China}

\author{Wei-Wei Yang}
\affiliation{Beijing National Laboratory for Condensed Matter Physics and Institute of Physics, Chinese Academy of Sciences, Beijing 100190, China}
\author{Xueming Guo}
\affiliation{Key Laboratory of Quantum Theory and Applications of MoE $\&$ School of Physical Science and Technology, Lanzhou University, Lanzhou 730000, People Republic of China}
    \affiliation{Lanzhou Center for Theoretical Physics, Key Laboratory of Theoretical Physics of Gansu Province, Lanzhou University, Lanzhou 730000, People Republic of China}

\author{Hong-Gang Luo}
\affiliation{Key Laboratory of Quantum Theory and Applications of MoE $\&$ School of Physical Science and Technology, Lanzhou University, Lanzhou 730000, People Republic of China}
\affiliation{Lanzhou Center for Theoretical Physics, Key Laboratory of Theoretical Physics of Gansu Province, Lanzhou University, Lanzhou 730000, People Republic of China}

\author{Yin Zhong}
\email{zhongy@lzu.edu.cn}
\affiliation{Key Laboratory of Quantum Theory and Applications of MoE $\&$ School of Physical Science and Technology, Lanzhou University, Lanzhou 730000, People Republic of China}
    \affiliation{Lanzhou Center for Theoretical Physics, Key Laboratory of Theoretical Physics of Gansu Province, Lanzhou University, Lanzhou 730000, People Republic of China}
\begin{abstract}
Recently, a novel collinear magnet, i.e. the altermagnet (AM), with spin-splitting energy band and zero net magnetization have attracted great interest due to its potential spintronic applications. Here, we demonstrate AM-like phases in a microscopic Kondo lattice (KL) model with an alternating next-nearest-neighbor-hopping (NNNH). Such alternating NNNH take nonmagnetic atoms, neglected in usual antiferromagnetism study, into account when encountering real-life candidate AM materials. With the framework of fermionic parton mean-field theory, we find three different ground-states for the half-filling KL: 1) a $d$-wave AM state; 2) a coexistent phase with both $d$-wave AM and intrinsic Kondo screening effect; 3) a Kondo insulator. The AM-like states are characterized by their spin-splitting quasiparticle bands, Fermi surface, spin-resolved distribution function and conductivity. It is suggested that the magnetic quantum oscillation, scanning tunneling microscopy and charge transport measurement can detect those AM-like phases. We hope the present work may be useful for exploring AM-like phases in $f$-electron compounds such as
CeNiAsO and Ce$_{4}$X$_{3}$(X=As,Sb,Bi).
\end{abstract}

\maketitle
\section{Introduction}
Recently, collinear magnetism beyond ferromagnetic and antiferromagnetic states, i.e. the altermagnetism (AM) with spin-splitting energy band and zero net magnetization have attracted great interest due to its potential spintronic applications.\cite{Mazin2022,Smejkal2022,Smejkal2022b,Bai2024} The AM state requires bipartite systems and the symmetry of combined time-reversal and translation is broken, which eliminates the
symmetry between time-reversal partners $|-k,\sigma\rangle$ and $|k,-\sigma\rangle$. Current candidate materials of AM are mainly $d$-electron compounds such as rutile metal RuO$_{2}$, chalcogenide semiconductor MnTe and pnictide metal CrSb and even the parent compound of high-$T_{c}$
cuprate superconductor La$_{2}$CuO$_{4}$,\cite{Krempasky2024,Fedchenko2024,Bai2022,Feng2022,Gonzalez2023,Bai2023,Hariki2024,Lee2024} wherein the smoking-gun feature of AM, namely the spin-dependent band-splitting with $d$ or $g$-wave symmetry has been unambiguously observed in RuO$_{2}$, MnTe and CrSb by angle-resolved photoemission spectroscopy (ARPES).\cite{Krempasky2024,Fedchenko2024,Lee2024,Osumi2024,Yang2024}

Although the density-functional-theory-based first-principle calculation has successfully predicted these mentioned $d$-electron AM materials,\cite{Smejkal2020,Mazin2021,Ahn2019,Yuan2020} more intuitive understanding, particularly the many-body physics involving topological superconductivity and Kondo effect,\cite{Chakraborty2023,Zhu2023,Diniz2024,Lee2023} can be extracted from microscopic lattice models.\cite{Hayami2019,Hayami2020,Brekke2023,Mland2024,Bose2024,Maier2023,Das2023,Leeb2024,Sato2024} In Ref.~\onlinecite{Das2023}, both metallic and insulating AM has been found in Hartree-Fock treatment of single-band Hubbard model with an alternating next-nearest-neighbor-hopping (NNNH), which could be realized in optical lattice. Physically, such alternating NNNH takes (nonequivalent) nonmagnetic atoms, neglected in usual antiferromagnetism study, into account when encountering real-life AM materials.\cite{Smejkal2022} This alternating NNNH combined with antiferromagnetic background from Hartree-Fock approximation reproduces the desirable AM phase. On the other hand, Ref.~\onlinecite{Brekke2023,Mland2024} underline the importance of the $s-d$ exchange model where itinerant electron interplays with local spin moment via Heisenberg/Hund-like exchange interaction. It is indeed that the antiferromagnetic configuration of local spin moment leads to the desirable AM state, in which fluctuations due to magnon-exchange drive the unconventional $p$-wave pairing.

Actually, such itinerant-local exchange interaction is ubiquitous in heavy fermion systems and the corresponding model turns to be the classic Kondo lattice model (KLM).\cite{Tsunetsugu,Coleman2015}
Considering the antiferromagnetic states have already occupied sensible parameter regime in the phase diagram of KLM,\cite{Lacroix1979,Fazekas1991,Zhang2000,Li2015,Watanabe2007,Asadzadeh2013,Martin2008,Lenz2008} it is promising to explore AM in heavy fermion compounds if effects of nonmagnetic atoms are suitably accounted for, e.g. by inserting the alternating NNNH into KLM.

\begin{figure}
\includegraphics[width=1.0\linewidth]{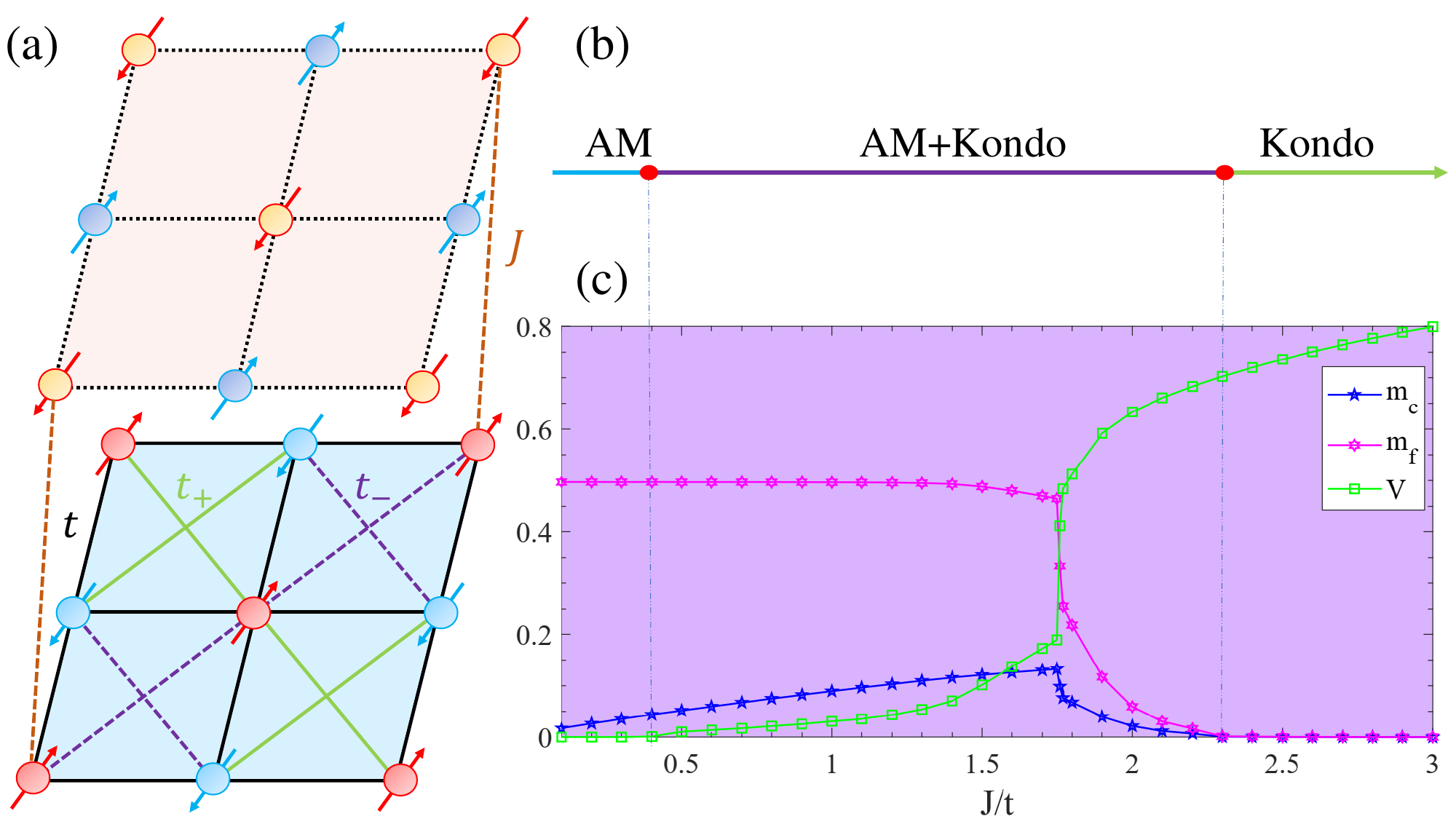}
\caption{\label{fig:1} (a) The Kondo lattice model (KLM) on square lattice with alternating next-nearest-neighbor-hopping (NNNH) $t_{+},t_{-}$. The upper layer denotes local spin moments, which interact with conduction electrons in the lower layer via Kondo coupling $J$. (b) The schematic ground-state phase diagram of KLM with alternating NNNH at half-filling. There exit three kinds of states, i.e. altermagnet (AM), Kondo state (metallic or insulator) and the coexistence of AM and Kondo state. (c) Mean-field order parameters $m_{c},m_{f},V$ evolve with increasing Kondo coupling $J$ in the ground-state of half-filled KLM. ($t=1,t_{+}=0.27,t_{-}=0.03,n_{c}=1,T=0$)}
\end{figure}

In this work, we take the first step towards addressing this issue and study a KLM with an alternating NNNH on square lattice. (Fig.~\ref{fig:1}(a)) With fermionic parton/slave-particle mean-field theory, we uncover $d$-wave AM with spin-splitting quasi-particle energy bands. Interestingly, although AM competes with Kondo screening, we find there exists a coexisting regime involving both AM and Kondo screening, which is reminiscent of the coexistence of antiferromagnetism and paramagnetic Kondo screening states in original KLM without alternating NNNH.\cite{Assaad1999,Zhang2000,Danu2021}

In addition to the spin-splitting bands, spin-resolved density of state, distribution function and conductivity are calculated and the latter two could be used to identify AM-like phases in generic heavy fermion compounds. For real-life heavy fermion materials, we suggest that the magnetic quantum oscillation, scanning tunneling microscopy and charge transport measurement can detect those AM-like phases. Going beyond the present mean-field theory tends to require sophisticated numerical techniques such as the constrained path Monte Carlo.\cite{SZhang1995} It will be interesting to perform such large-scale simulation in future work and more importantly, we hope the present work is able to stimulate exploration of AM candidates in realistic heavy fermion materials.

The remaining part of this article is organized as follows. In Sec.~\ref{sec1}, we introduce the KLM on square lattice with an alternating NNNH. Formalism of fermionic parton mean-field theory is presented with details. Sec.~\ref{sec2} provides the solution of mean-field theory and possible phases including the Kondo screening state, AM state and their coexistent state are analyzed. Physical observable like the spin-resolved density of state and conductivity are considered. Sec.~\ref{sec3} is devoted to some discussions, e.g. the implication for experiment and realistic
materials and alternative theoretical approaches. Brief summary is given in Sec.~\ref{sec4}.

\section{Model and mean-field theory}\label{sec1}
Let us consider the following KLM on square lattice, (extension to other lattice geometries is straightforward) which has been believed to be the standard model for heavy fermion compounds\cite{Tsunetsugu,Coleman2015}
\begin{equation}
\hat{H}=-\sum_{ij,\sigma}t_{ij}\hat{c}_{i\sigma}^{\dag}\hat{c}_{j\sigma}+\frac{J}{2}\sum_{j,\sigma\sigma'}\hat{c}_{j\sigma}^{\dag}\vec{\sigma}_{\sigma\sigma'}\hat{c}_{j\sigma'}\cdot \hat{\vec{S}}_{j}\label{eq1}
\end{equation}
Here, $\hat{c}_{j\sigma}^{\dag}$ is the creation operator of conduction electron ($c$-electron) with spin flavor $\sigma=\uparrow,\downarrow$ while $\hat{\vec{S}}_{j}$ denotes local spin moment of $f$-electron at site $j$. The $c$-electron obeys the usual anti-commutative rules, i.e. $\{\hat{c}_{i\sigma},\hat{c}_{j\sigma'}^{\dag}\}=
\delta_{ij}\delta_{\sigma\sigma'}$, $\{\hat{c}_{i\sigma}^{\dag},\hat{c}_{j\sigma'}^{\dag}\}=\{\hat{c}_{i\sigma},\hat{c}_{j\sigma'}\}=0$. The local moment $\hat{\vec{S}}_{j}$ is encoded by the Pauli matrix $\frac{1}{2}\hat{\vec{\sigma}}_{j}$ associated with $[\hat{\sigma}_{m}^{\alpha},\hat{\sigma}_{n}^{\beta}]=2i\epsilon^{\alpha\beta\gamma}\hat{\sigma}^{\gamma}_{m}\delta_{mn}$.

$t_{ij}$ are hopping integral between $i,j$ sites for $c$-electron. Inspired by Ref.~\onlinecite{Das2023}, we consider nearest-neighbor-hopping (NNH) $t$ and NNNH $t_{+},t_{-}$ so as to obtain AM in KLM. (The choice of the latter ones is shown in Fig.~\ref{fig:1}(a), where the green (purple) line corresponds to $t_{+}$ ($t_{-}$).) The appearance of alternating NNNH breaks the $\pi/2$-rotation ($C_{4z}$) symmetry of the original square lattice and this feature is essential for producing AM phase. Physically, such alternating NNNH takes (nonequivalent) nonmagnetic atoms, neglected in usual antiferromagnetism study, into account when encountering real-life AM materials.\cite{Smejkal2022} Thus, AM materials may be explored in existing antiferromagnets with symmetry-breaking such as the mentioned $\pi/2$-rotation.

$J$ denotes the strength of Kondo coupling between $c$-electron and local spin. Furthermore, chemical potential $\mu$ can be added to fix electron's density $n_{c}$. It is noted that, when NNNH does not vanish, the KLM cannot be solved by unbiased quantum Monte Carlo simulation even in half-filling case (density of $c$-electron $n_{c}=1$) due to the fermion sign problem.\cite{Assaad1999}

For $t_{+}\neq t_{-}$, the lattice is divided into $A$ and $B$ sublattice, thus we can rewrite the Hamiltonian as follows,
\begin{equation}
\hat{H}=\hat{H}_{0}+\hat{H}_{K},
\end{equation}
where the non-interacting part reads
\begin{eqnarray}
\hat{H}_{0}&=&-t\sum_{i,\delta,\sigma}(\hat{c}_{iA\sigma}^{\dag}\hat{c}_{i+\delta,B\sigma}+\hat{c}_{i+\delta,B\sigma}^{\dag}\hat{c}_{iA\sigma})\nonumber\\
&-&\sum_{i,\delta_{1}',\sigma}(t_{+}\hat{c}_{iA\sigma}^{\dag}\hat{c}_{i+\delta_{1}',A\sigma}+t_{-}\hat{c}_{iB\sigma}^{\dag}\hat{c}_{i+\delta_{1}',B\sigma})\nonumber\\
&-&\sum_{i,\delta_{2}',\sigma}(t_{-}\hat{c}_{iA\sigma}^{\dag}\hat{c}_{i+\delta_{2}',A\sigma}+t_{+}\hat{c}_{iB\sigma}^{\dag}\hat{c}_{i+\delta_{2}',B\sigma}).\label{eq4}
\end{eqnarray}
The $\delta=(\pm1,0),(0,\pm1)$ represents the vector of NNH. As for the vector of NNNH, we have
$\delta_{1}'=(1,1),(-1,-1)$ and $\delta_{2}'=(1,-1),(-1,1)$. (Fig.~\ref{fig:1}(a))

\begin{eqnarray}
\hat{H}_{K}&=&\frac{J}{2}\sum_{j,\alpha=A,B,\sigma\sigma'}\hat{c}_{j\alpha\sigma}^{\dag}\vec{\sigma}_{\sigma\sigma'}\hat{c}_{j\alpha\sigma'}\cdot \hat{\vec{S}}_{j\alpha}\nonumber\\
&=&\frac{J}{2}\sum_{j}(\hat{c}_{jA\uparrow}^{\dag}\hat{c}_{jA\downarrow}\hat{S}_{jA}^{-}+\hat{c}_{jA\downarrow}^{\dag}\hat{c}_{jA\uparrow}\hat{S}_{jA}^{+})\nonumber\\
&+&\frac{J}{2}\sum_{j}(\hat{c}_{jB\uparrow}^{\dag}\hat{c}_{jB\downarrow}\hat{S}_{jB}^{-}+\hat{c}_{jB\downarrow}^{\dag}\hat{c}_{jB\uparrow}\hat{S}_{jB}^{+})\nonumber\\
&+&\frac{J}{2}\sum_{j,\sigma}\sigma(\hat{c}_{jA\sigma}^{\dag}\hat{c}_{jA\sigma}\hat{S}_{jA}^{z}+\hat{c}_{jB\sigma}^{\dag}\hat{c}_{jB\sigma}\hat{S}_{jB}^{z}).
\end{eqnarray}
Here, the first two lines correspond to the transverse Kondo coupling, which gives rise to Kondo screening in the mean-field level. In contrast, the longitudinal Kondo coupling in the last line is responsible for the appearance of magnetic long-ranged order. Such magnetic orders compete with Kondo screening and would lead to magnetic-paramagnetic transition.

Now, it is straightforward to have $\hat{H}_{0}$ in momentum space in terms of Fourier transformation,
\begin{eqnarray}
\hat{H}_{0}&=&\sum_{k\sigma}\left(
                   \begin{array}{cc}
                     \hat{c}_{kA\sigma}^{\dag} & \hat{c}_{kB\sigma}^{\dag} \\
                   \end{array}
                 \right)\left(
                          \begin{array}{cc}
                            \varepsilon_{k}^{AA} & \varepsilon_{k} \\
                            \varepsilon_{k} & \varepsilon_{k}^{BB} \\
                          \end{array}
                        \right)
\left(
                   \begin{array}{c}
                     \hat{c}_{kA\sigma} \\
                     \hat{c}_{kB\sigma} \\
                   \end{array}
                 \right)
\end{eqnarray}
and
\begin{eqnarray}
&&\varepsilon_{k}=-2t(\cos k_{x}+\cos k_{y})\nonumber\\
&&\varepsilon_{k}^{AA}=-2t_{+}\cos(k_{x}+k_{y})-2t_{-}\cos(k_{x}-k_{y})\nonumber\\
&&\varepsilon_{k}^{BB}=-2t_{-}\cos(k_{x}+k_{y})-2t_{+}\cos(k_{x}-k_{y})\nonumber.
\end{eqnarray}
Here, $\varepsilon_{k},\varepsilon_{k}^{AA},\varepsilon_{k}^{BB}$ denote intra-sublattice and inter-sublattice hopping energy, respectively. Since one has chosen $A,B$ sublattice structure, the Brillouin zone (BZ) used for summation over momentum $k$ must be the magnetic BZ or reduced BZ, which is determined by $\cos k_{x}+\cos k_{y}\geqslant0$. When $t_{+}=t_{-}=t_{1}$, $\varepsilon_{k}^{AA}=\varepsilon_{k}^{BB}=-4t_{1}\cos k_{x}\cos k_{y}$, which is just the standard result for an isotropic NNNH term on square lattice.

Although the Kondo interaction term $\hat{H}_{K}$ has simple structure in real space (Hilbert space has $8$-dimension on each site), it cannot be diagonalized via Fourier transformation, in contrast to the case of infinite-ranged Hatsugai-Kohmoto-type interaction.\cite{Zhong2022,Wang2024} It is well-known that treating $\hat{H}_{K}$ and $\hat{H}_{0}$ on the equal footing is still challenging in terms of current analytical or numerical techniques,\cite{Tsunetsugu,Watanabe2007,Asadzadeh2013,Martin2008,Lenz2008} so we try the well-developed fermionic parton mean-field theory.\cite{Coleman2015,Zhang2000,Li2015,Hewson1993}

In practise, the $S=1/2$ local spin moment is splitting into two flavor fermions as
\begin{equation}
\hat{\vec{S}}_{j}=\frac{1}{2}\sum_{\sigma\sigma'}\hat{f}^{\dag}_{j\sigma}\vec{\sigma}_{\sigma\sigma'}\hat{f}_{j\sigma'}
\end{equation}
where $\hat{f}_{j\sigma}$ is the Abriksov/Schwinger fermion operator and satisfies the anticommutative relation $\{\hat{f}_{i\sigma},\hat{f}_{j\sigma'}^{\dag}\}=\delta_{ij}\delta_{\sigma\sigma'}$. The above process is an example of fractonalization because $\hat{f}_{j\sigma}$ carries $S=1/2$ degree of freedom instead of magnon (the $S=1$ excitation). Thus, this fermion is also called spinon in the literature of quantum spin liquid and high-$T_{c}$ cuprate superconductor.\cite{Zhou2017,Lee2006} Note that the local spin moment $\hat{\vec{S}}_{j}$ is invariant under the $U(1)$ gauge transformation $\hat{f}_{j\sigma}\rightarrow e^{i\theta_{j}}\hat{f}_{j\sigma}$. When Kondo screening is active, the $U(1)$ gauge symmetry is breaking and $\hat{f}_{j\sigma}$ acquires electric charge via Higgs mechanism. In this situation, the Abriksov fermion can be seen as a composite fermion $\hat{f}_{j\sigma}\sim \hat{c}_{j\sigma'}\vec{\sigma}_{\sigma'\sigma}\cdot\hat{\vec{S}}_{j}$.\cite{Coleman2015,Chen2024}

One must be careful to apply above fermionic representation since those fermions should satisfy the constraint $\sum_{\sigma}\hat{f}^{\dag}_{j\sigma}\hat{f}_{j\sigma}=1$ at each site to exclude nonphysical Hilbert space. Violation of this constraint may lead to nonphysical results even for the simplest case of Hubbard dimer.\cite{WWYang2019} In practice, this constraint can only be treated approximately for the mean-field theory given below. It seems that this treatment may eliminate the variational control associated with the mean-field theory and lead to an uncontrolled Hartree-Fock-like calculation.

After utilizing the Abriksov fermion representation, the Kondo coupling term is rewritten as
\begin{eqnarray}
\hat{H}_{K}&=&\frac{J}{2}\sum_{j}(\hat{c}_{jA\uparrow}^{\dag}\hat{c}_{jA\downarrow}\hat{f}_{jA\downarrow}^{\dag}\hat{f}_{jA\uparrow}+\hat{c}_{jA\downarrow}^{\dag}\hat{c}_{jA\uparrow}\hat{f}_{jA\uparrow}^{\dag}\hat{f}_{jA\downarrow})\nonumber\\
&+&\frac{J}{2}\sum_{j}(\hat{c}_{jB\uparrow}^{\dag}\hat{c}_{jB\downarrow}\hat{f}_{jB\downarrow}^{\dag}\hat{f}_{jB\uparrow}+\hat{c}_{jB\downarrow}^{\dag}\hat{c}_{jB\uparrow}\hat{f}_{jB\uparrow}^{\dag}\hat{f}_{jB\downarrow})\nonumber\\
&+&\frac{J}{4}\sum_{j,\sigma\sigma'}\sigma\sigma'(\hat{c}_{jA\sigma}^{\dag}\hat{c}_{jA\sigma}\hat{f}_{jA\sigma'}^{\dag}\hat{f}_{jA\sigma'}+\hat{c}_{jB\sigma}^{\dag}\hat{c}_{jB\sigma}\hat{f}_{jB\sigma'}^{\dag}\hat{f}_{jB\sigma'})\nonumber\\
&&\label{eq5}
\end{eqnarray}
Because, we will use mean-field approximation to treat $\hat{H}_{K}$, it is helpful to introduce Kondo screening order parameters
\begin{equation}
V=-\langle \hat{c}^{\dag}_{j\alpha\sigma}\hat{f}_{j\alpha\sigma}\rangle=-\langle \hat{f}^{\dag}_{j\alpha\sigma}\hat{c}_{j\alpha\sigma}\rangle,~\alpha=A,B\nonumber
\end{equation}
and magnetic order parameters
\begin{eqnarray}
&&m_{A}^{c}=\frac{1}{2}\sum_{\sigma}\sigma\langle \hat{c}_{jA\sigma}^{\dag}\hat{c}_{jA\sigma}\rangle,m_{A}^{f}=\frac{1}{2}\sum_{\sigma}\sigma\langle \hat{f}_{jA\sigma}^{\dag}\hat{f}_{jA\sigma}\rangle,\nonumber\\
&&m_{B}^{c}=\frac{1}{2}\sum_{\sigma}\sigma\langle \hat{c}_{jB\sigma}^{\dag}\hat{c}_{jB\sigma}\rangle,m_{B}^{f}=\frac{1}{2}\sum_{\sigma}\sigma\langle \hat{f}_{jB\sigma}^{\dag}\hat{f}_{jB\sigma}\rangle.\nonumber
\end{eqnarray}
A careful reader may note that in the conventional large-N approach for Kondo lattice, the magnetic channel is not included and the magnetic order parameters are absence.\cite{Coleman2015} As discussed in previous works on AM, one has to assume antiferromagnetic magnetic structure to generate AM,\cite{Brekke2023,Mland2024,Bose2024,Maier2023,Das2023,Leeb2024} which motivates us to consider
\begin{equation}
m_{A}^{c}=-m_{B}^{c}=m_{c},~~m_{A}^{f}=-m_{B}^{f}=-m_{f}.
\end{equation}
This choice leads to the collinear antiferromagnetic configuration shown in Fig.~\ref{fig:1}, where $c$-electron and local spin on the same site have the opposite spin direction, consisting with the antiferromagnetic Kondo coupling. (If this coupling is ferromagnetic as in the Hund interaction, one may assume $m_{A}^{c}=-m_{B}^{c}=m_{c},~~m_{A}^{f}=-m_{B}^{f}=m_{f}$. In Appendix.~\ref{Ap5}, we have explored the possibility of stripe order, which competes with the antiferromagnetism. It is found that if the strength of NNNH is smaller than NNH, the antiferromagnetic configuration is still stable.) Under this antiferromagnetic configuration, the system is invariant under the combination operation (time-reversal, $C_{4z}$ and translation), which is denoted by $[C_{2}\|C_{4z}t]$ (the spin-group symmetry) and it gives rise to the so-called $d$-wave AM.\cite{Smejkal2022} As comparison, the conventional antiferromagnetism is invariant under the time-reversal combined with translation (denoting $[C_{2}\|t]$) or just the single operation $C_{4z}$. In this sense, AM may be considered as a secondary symmetry-breaking state of the usual antiferromagnetic state.

Then, using the standard mean-field decoupling $\hat{A}\hat{B}\simeq\langle \hat{A}\rangle \hat{B}+\hat{A}\langle \hat{B}\rangle-\langle \hat{A}\rangle\langle \hat{B}\rangle$ and Fourier transformation, we obtain
\begin{eqnarray}
\hat{H}_{K}&\simeq&\frac{J}{2}V\sum_{k\sigma}(\hat{c}_{kA\sigma}^{\dag}\hat{f}_{kA\sigma}+\hat{f}_{kA\sigma}^{\dag}\hat{c}_{kA\sigma})\nonumber\\
&+&\frac{J}{2}V\sum_{k\sigma}(\hat{c}_{kB\sigma}^{\dag}\hat{f}_{kB\sigma}+\hat{f}_{kB\sigma}^{\dag}\hat{c}_{kB\sigma})\nonumber\\
&+&\frac{J}{2}\sum_{k\sigma}\sigma(m_{c}\hat{f}_{kA\sigma}^{\dag}\hat{f}_{kA\sigma}-m_{f}\hat{c}_{kA\sigma}^{\dag}\hat{c}_{kA\sigma})\nonumber\\
&+&\frac{J}{2}\sum_{k\sigma}\sigma(-m_{c}\hat{f}_{kB\sigma}^{\dag}\hat{f}_{kB\sigma}+m_{f}\hat{c}_{kB\sigma}^{\dag}\hat{c}_{kB\sigma})\nonumber\\
&+&2N_{s}J(V^{2}+m_{c}m_{f}).
\end{eqnarray}
where $N_{s}$ is the number of unit-cell and the number of sites should be $2N_{s}$.

Now, combining $\hat{H}_{0}$ with $\hat{H}_{K}$, we find the mean-field Hamiltonian of KLM as
\begin{equation}
\hat{H}=\sum_{k\sigma}\hat{\psi}_{k\sigma}^{\dag}H_{\sigma}(k)\hat{\psi}_{k\sigma}+2N_{s}(JV^{2}+Jm_{c}m_{f}-\lambda)
\end{equation}
with the spinor $\hat{\psi}_{k\sigma}^{\dag}=(\hat{c}_{kA\sigma}^{\dag},\hat{c}_{kB\sigma}^{\dag},\hat{f}_{kA\sigma}^{\dag},\hat{f}_{kB\sigma}^{\dag})$ and $4\times4$-matrix
\begin{equation}
H_{\sigma}(k)=\left(
                \begin{array}{cc}
                  H_{\sigma}^{cc}(k) & H_{\sigma}^{cf}(k) \\
                  H_{\sigma}^{fc}(k) & H_{\sigma}^{ff}(k) \\
                \end{array}
              \right)
\end{equation}
\begin{equation}
H_{\sigma}^{cc}(k)=\left(
                \begin{array}{cc}
                  \varepsilon_{k}^{AA}-\mu-\frac{Jm_{f}\sigma}{2} & \varepsilon_{k} \\
                  \varepsilon_{k} & \varepsilon_{k}^{BB}-\mu+\frac{Jm_{f}\sigma}{2} \\
                \end{array}
              \right)
\end{equation}
\begin{equation}
H_{\sigma}^{ff}(k)=\left(
                \begin{array}{cc}
                  \lambda+\frac{Jm_{c}\sigma}{2} & 0 \\
                  0 & \lambda-\frac{Jm_{c}\sigma}{2} \\
                \end{array}
              \right)
\end{equation}
\begin{equation}
H_{\sigma}^{cf}(k)=H_{\sigma}^{fc}(k)=\left(
                \begin{array}{cc}
                  \frac{JV}{2} & 0 \\
                  0 & \frac{JV}{2} \\
                \end{array}
              \right)
\end{equation}
Here, chemical potential $\mu$ and Lagrangian multiplier $\lambda$ are included, whose effects are fixing conduction electron density $n_{c}$ and enforcing the constraint on average,
\begin{equation}
\frac{1}{2N_{s}}\sum_{k\sigma}(\langle \hat{c}_{kA\sigma}^{\dag}\hat{c}_{kA\sigma}\rangle+\langle \hat{c}_{kB\sigma}^{\dag}\hat{c}_{kB\sigma}\rangle)=n_{c} \label{eq:MF1}
\end{equation}
and
\begin{equation}
\frac{1}{2N_{s}}\sum_{k\sigma}(\langle \hat{f}_{kA\sigma}^{\dag}\hat{f}_{kA\sigma}\rangle+\langle \hat{f}_{kB\sigma}^{\dag}\hat{f}_{kB\sigma}\rangle)=1. \label{eq:MF2}
\end{equation}

Unfortunately, the $4\times4$-matrix $H_{\sigma}(k)$ have complicated algebraic structures for its eigenstates and eigenvalues, thus it is not possible to derive mean-field self-consistent equations required to give order parameters $V,m_{c}$ and $m_{f}$. However, it is helpful to use the definition of $V,m_{c},m_{f}$, which gives rise to
\begin{equation}
V=-\frac{1}{N_{s}}\sum_{k}\langle \hat{c}^{\dag}_{kA\sigma}\hat{f}_{kA\sigma}\rangle,\label{eq:MF3}
\end{equation}
\begin{equation}
m_{c}=\frac{1}{2N_{s}}\sum_{k\sigma}\sigma\langle \hat{c}^{\dag}_{kA\sigma}\hat{c}_{kA\sigma}\rangle,\label{eq:MF4}
\end{equation}
\begin{equation}
m_{f}=-\frac{1}{2N_{s}}\sum_{k\sigma}\sigma\langle \hat{f}^{\dag}_{kA\sigma}\hat{f}_{kA\sigma}\rangle.\label{eq:MF5}
\end{equation}
Thus, utilizing Eqs.~\ref{eq:MF3},\ref{eq:MF4},\ref{eq:MF5} combining with Eqs.~\ref{eq:MF1} and \ref{eq:MF2}, we are able to find all order parameters, chemical potential and Lagrangian multiplier via
the standard iterative algorithm supplemented with the bisection method.\cite{Li2015}

\section{Results of mean-field theory}\label{sec2}
Guided by previous works on square lattice KLM without alternating NNNH at half-filling,\cite{Assaad1999,Zhang2000,Li2015,Danu2021,Capponi2001} we first explore the ground-state phase diagram of KLM with alternating NNNH at half-filling ($n_{c}=1$). (We have used $t_{+}=0.57,t_{-}=0.03$ in the main text and examples with other $t_{+},t_{-}$ have been discussed in Appendix.~\ref{Ap1}.)

According to the Kondo screening order parameter $V$, magnetic order parameters $m_{c}$ and $m_{f}$, we expect four kinds of ground-states:
\begin{enumerate}
  \item Trivial state with $V=m_{c}=m_{f}=0$;
  \item Kondo screening state with $V\neq0$ and $m_{c}=m_{f}=0$, it dominates at large $J$;
  \item Magnetic state with $V=0$ and $m_{c},m_{f}\neq0$, which is stable for small $J$ regime;
  \item Coexistent state with all $V,m_{c},m_{f}$ being nonzero. It occupies the phase diagram in the intermediate $J$ regime.
\end{enumerate}
Except for the Kondo state and the decoupled trivial state, we will see below that both the magnetic state and the coexistent state can be identified as AM due to their spin-splitting quasiparticle energy bands. To emphasize the coexistence with Kondo screening, we say there is a coexistent phase with AM and Kondo screening. (Fig.~\ref{fig:1}(b) and (c)) This coexistent phase is not contradicted Landau's phase transition theory since orders in AM is the antiferromagnetic order and Kondo screening state is a Higgs phase of the underlying lattice gauge theory.\cite{Coleman2015}
\subsection{Trivial state}
When all mean-field parameters $V,m_{c},m_{f}$ are zero, there is only a trivial state, which is described by the Hamiltonian,
\begin{eqnarray}
\hat{H}=\sum_{k\sigma}\left(
                   \begin{array}{cc}
                     \hat{c}_{kA\sigma}^{\dag} & \hat{c}_{kB\sigma}^{\dag} \\
                   \end{array}
                 \right)\left(
                          \begin{array}{cc}
                            \varepsilon_{k}^{AA}-\mu & \varepsilon_{k} \\
                            \varepsilon_{k} & \varepsilon_{k}^{BB}-\mu \\
                          \end{array}
                        \right)
\left(
                   \begin{array}{c}
                     \hat{c}_{kA\sigma} \\
                     \hat{c}_{kB\sigma} \\
                   \end{array}
                 \right).\nonumber
\end{eqnarray}
Its dispersion is
\begin{equation}
E_{k\sigma\pm}=\frac{1}{2}(\varepsilon_{k}^{AA}+\varepsilon_{k}^{BB}\pm\sqrt{(\varepsilon_{k}^{AA}-\varepsilon_{k}^{BB})^{2}+4\varepsilon_{k}^{2}})-\mu,\nonumber
\end{equation}
which tells us that the Kondo coupling does not work at the mean-field level. Actually, fluctuations of the Kondo interaction term must exist and it may be included in terms of perturbation theory or treated by equation of motion method for the retarded Green's function.\cite{Bernhard1999} At the present mean-field level, this decoupled trivial state is unstable to other ordered states and we will not consider it hereafter.
\subsection{Kondo screening state}
The Kondo screening state is rather simple to understand and it is the dominating phase at strong coupling ($J$ is large). In this situation, $m_{c}=m_{f}=0$, and this corresponds to the usual large-$N$ solution of Kondo lattice.\cite{Coleman2015} It is safe to turn off $t_{+},t_{-}$ since no AM is involved. Thus, the half-filling case has the particle-hole symmetry, which dictates $\mu=\lambda=0$. Consequently, the mean-field Hamiltonian reads,
\begin{equation}
\hat{H}=\sum_{k\sigma}\left(\varepsilon_{k}\hat{c}^{\dag}_{k\sigma}\hat{c}_{k\sigma}+\frac{JV}{2}\hat{c}^{\dag}_{k\sigma}\hat{f}_{k\sigma}+\frac{JV}{2}\hat{f}^{\dag}_{k\sigma}\hat{c}_{k\sigma}\right).
\end{equation}
Careful reader may note that the sublattice index $A,B$ does not appear as a result of vanishing $t_{+},t_{-}$. The key feature of the above Hamiltonian is that it gives two spin-degenerated quasi-particle bands
\begin{equation}
E_{k\sigma\pm}=\frac{1}{2}\left(\varepsilon_{k}\pm\sqrt{\varepsilon_{k}^{2}+J^{2}V^{2}}\right).
\end{equation}
At half-filling, the lower band $E_{k\sigma-}$ is fully occupied and the system belongs to be an insulator, which is usually called Kondo insulator in the language of heavy fermions.\cite{Coleman2015,Riseborough2000} This interaction-driven insulator has direct gap $JV$ and the indirect gap is about $J^{2}V^{2}/W$. ($W$ is the band-width of conduction electron) Alternatively, the physics of Kondo insulator can be captured by the strong coupling expansion, where the product of local spin singlet forms the basis and it gives the same qualitative result with the present mean-field theory.\cite{Coleman2015,Chen2024,Trebst2006,Huecker2023}

When we turn on $t_{+},t_{-}$, the mean-field Hamiltonian reads
\begin{equation}
\hat{H}=\sum_{k\sigma}\hat{\psi}_{k\sigma}^{\dag}H_{\sigma}(k)\hat{\psi}_{k\sigma}+2N_{s}(JV^{2}-\lambda)
\end{equation}
and
\begin{equation}
H_{\sigma}(k)=\left(
                \begin{array}{cccc}
                  \varepsilon_{k}^{AA}-\mu & \varepsilon_{k} & \frac{JV}{2} & 0 \\
                  \varepsilon_{k} & \varepsilon_{k}^{BB}-\mu & 0 & \frac{JV}{2} \\
                  \frac{JV}{2} & 0 & \lambda & 0 \\
                  0 & \frac{JV}{2} & 0 & \lambda \\
                \end{array}
              \right).
\end{equation}
It is clear that quantities in $H_{\sigma}(k)$ do not have dependence on $\sigma$, thus $\hat{H}$ only has spin-degenerated bands $E_{k\uparrow}=E_{k\downarrow}$, agreeing with the case of $t_{+}=t_{-}=0$.
\begin{figure}
\includegraphics[width=1.0\linewidth]{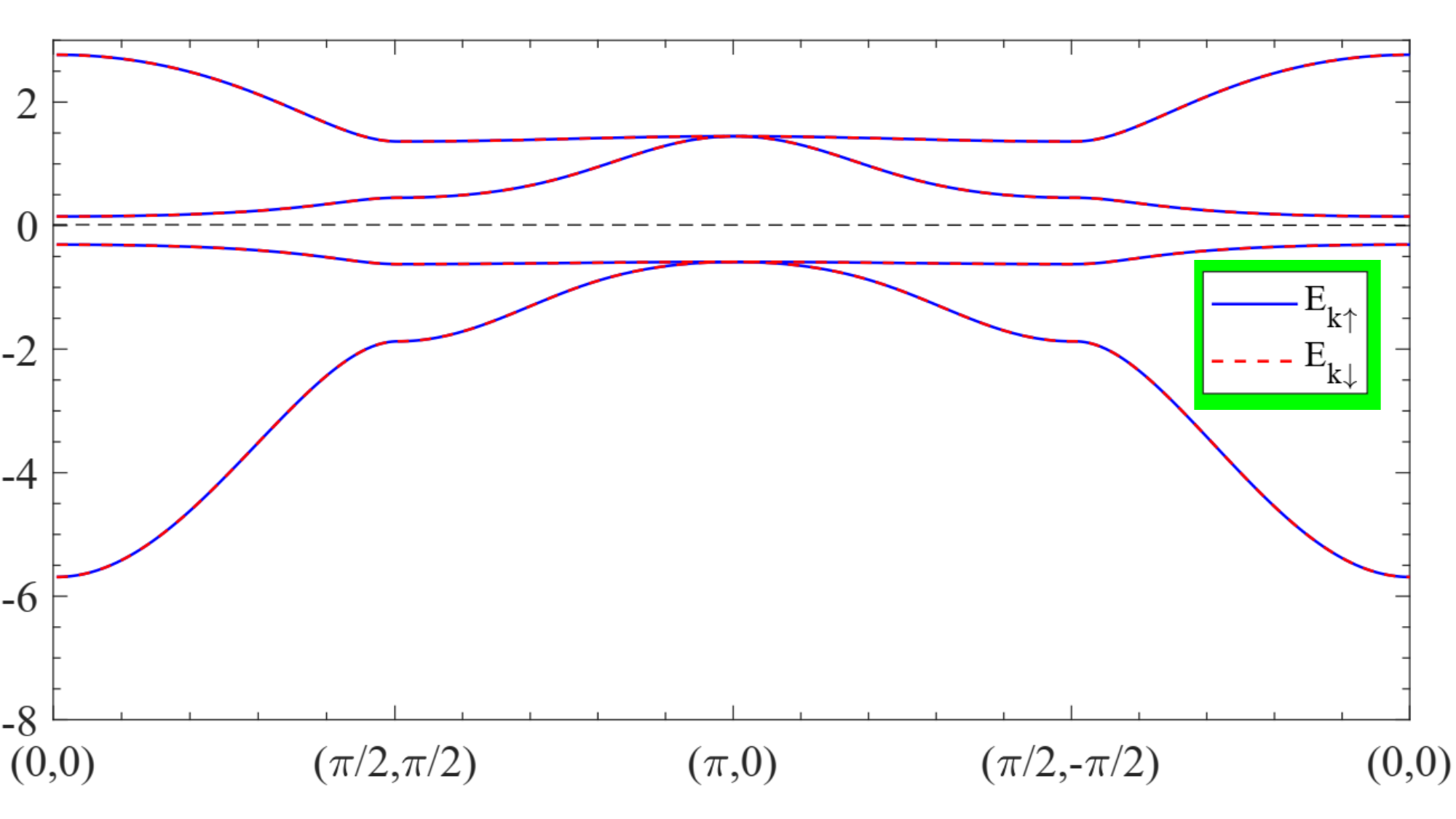}
\caption{\label{fig:3} The spin-degenerated bands in the Kondo screening state of KLM. This Kondo screening state corresponds to Kondo insulator with indirect gap $J^{2}V^{2}/W$. ($t=1,t_{+}=0.57,t_{-}=0.03,J=2.5,n_{c}=1,T=0$)}
\end{figure}

The example in Fig.~\ref{fig:3} confirms the above reasoning that the Kondo screening state of KLM only has
spin-degenerated bands. The indirect gap from Fig.~\ref{fig:3} is about $0.45$, which is consistent with expression $J^{2}V^{2}/W$ if we note that $V=0.736,J=2.5,W\simeq8$. Another noticeable feature is that the system at half-filling is always a Kondo insulator in spite of nonzero alternating NNNH. This fact seems to be valid for generic parameters if the Kondo screening solution is stable.
\subsection{AM state}
\begin{figure}
\includegraphics[width=1.0\linewidth]{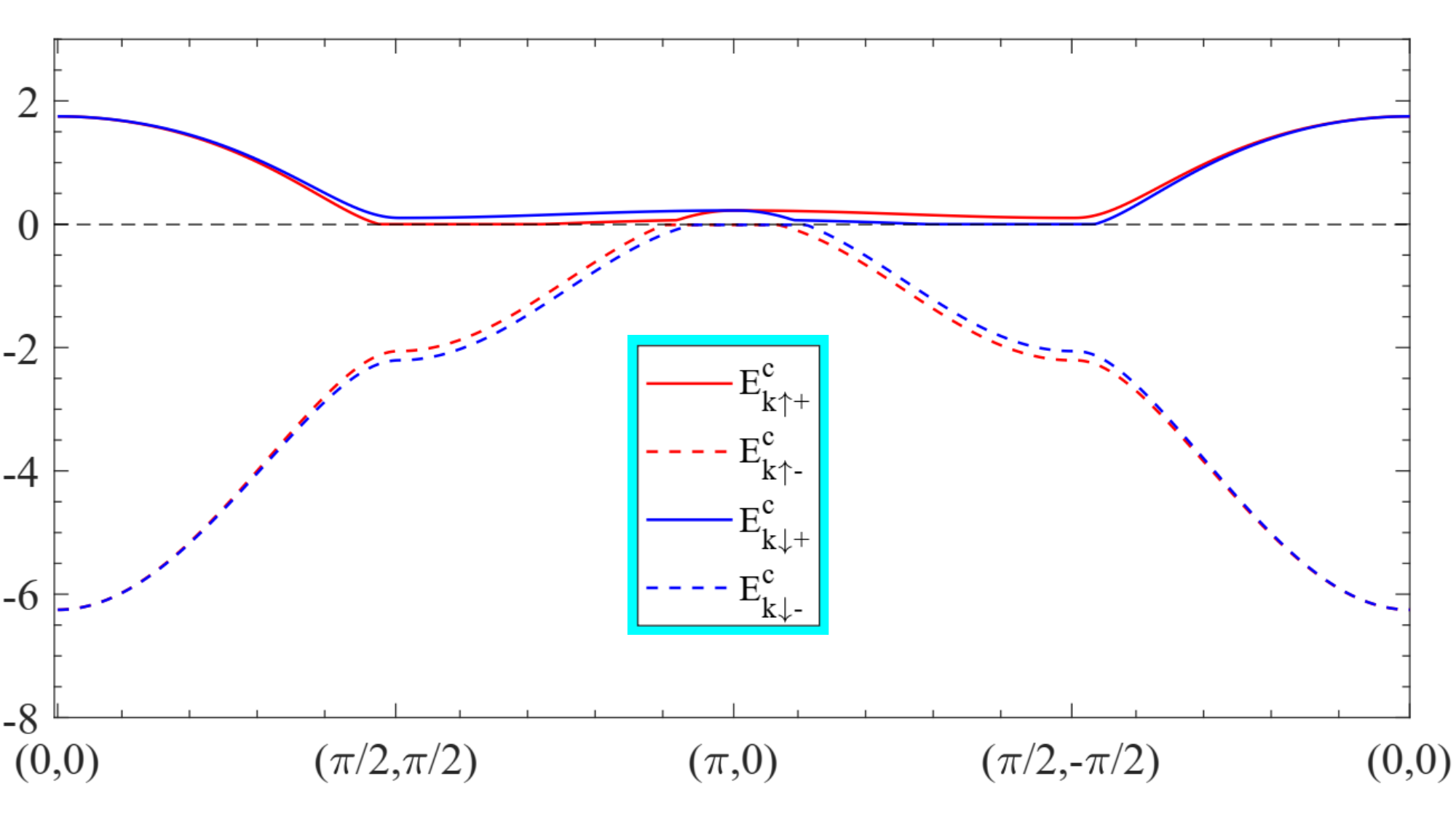}
\caption{\label{fig:2} The spin-splitting bands in the AM state of KLM. The $C_{4z}$ symmetry of bands implies a $d$-wave AM. ($t=1,t_{+}=0.57,t_{-}=0.03,J=0.3,n_{c}=1,T=0$)}
\end{figure}

\begin{figure}
\includegraphics[width=1.0\linewidth]{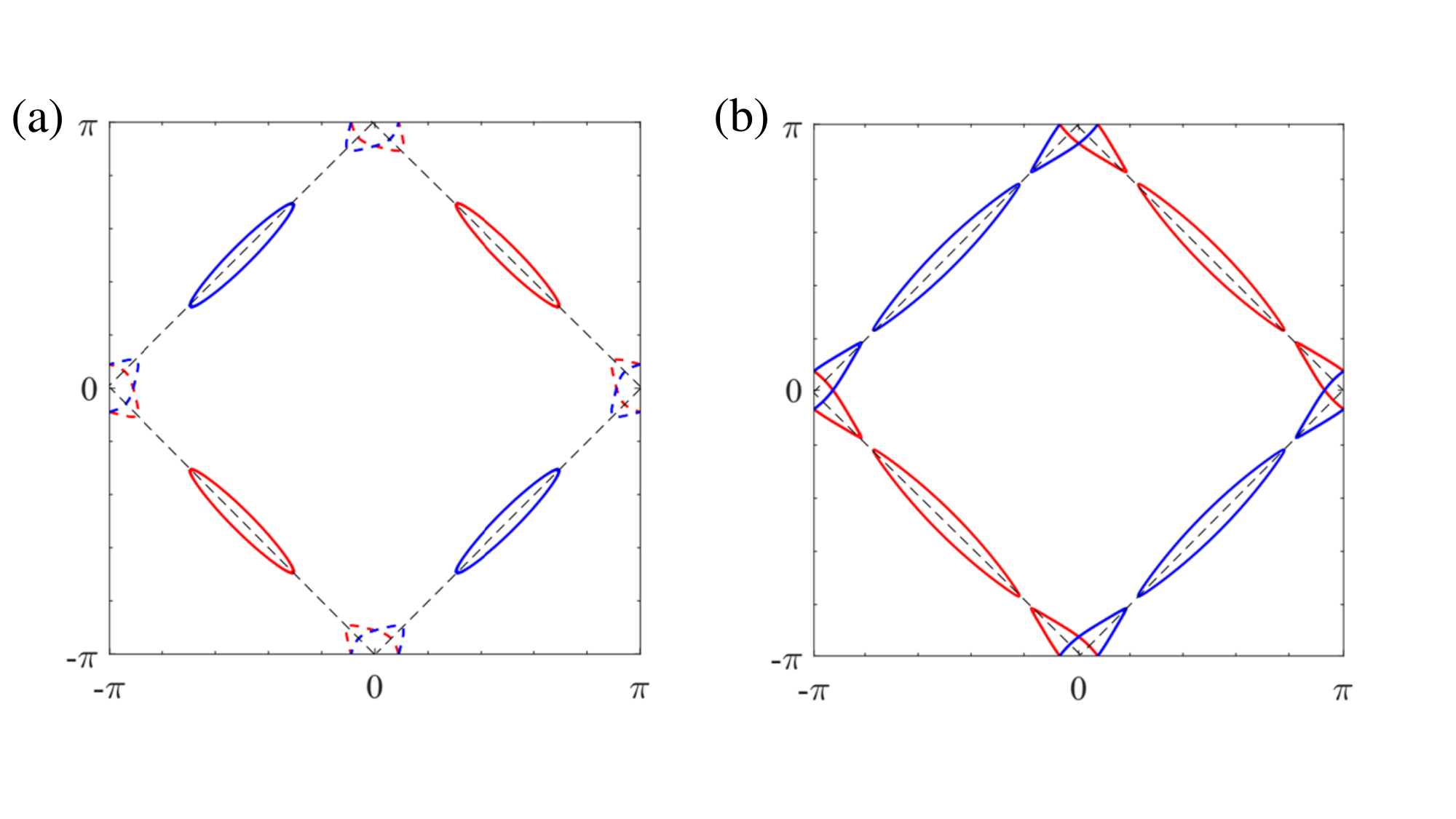}
\caption{\label{fig:22} The spin-splitting Fermi surface in (a) the AM with $J=0.3$ and (b) the coexistent states with $J=1.5$ of KLM. The $C_{4z}$ symmetry of Fermi surface implies a $d$-wave AM. ($t=1,t_{+}=0.57,t_{-}=0.03,n_{c}=1,T=0$)}
\end{figure}

In the pure AM state without Kondo screening, the conduction electron and the local spin are decoupled approximately, so the mean-field Hamiltonian is simplified into
\begin{eqnarray}
\hat{H}&=&\sum_{k\sigma}\left(
                        \begin{array}{cc}
                          \hat{c}^{\dag}_{kA\sigma} & \hat{c}^{\dag}_{kB\sigma} \\
                        \end{array}
                      \right)
H_{\sigma}^{cc}(k)\left(
                        \begin{array}{c}
                          \hat{c}_{kA\sigma} \\
                          \hat{c}_{kB\sigma} \\
                        \end{array}
                      \right)\nonumber\\
&+&\sum_{k\sigma}\left(
                        \begin{array}{cc}
                          \hat{f}^{\dag}_{kA\sigma} & \hat{f}^{\dag}_{kB\sigma} \\
                        \end{array}
                      \right)
H_{\sigma}^{ff}(k)\left(
                        \begin{array}{c}
                          \hat{f}_{kA\sigma} \\
                          \hat{f}_{kB\sigma} \\
                        \end{array}
                      \right).\label{eq2}
\end{eqnarray}
It is straightforward to find the following four quasiparticle bands,
\begin{eqnarray}
E_{k\sigma\pm}^{f}&=&\lambda\pm\frac{Jm_{c}\sigma}{2},\nonumber\\
E_{k\sigma\pm}^{c}&=&\frac{1}{2}\left(\varepsilon^{AA}_{k}+\varepsilon^{BB}_{k}\pm\sqrt{(\varepsilon_{k}^{AA}-\varepsilon_{k}^{BB}-Jm_{f}\sigma)^{2}+4\varepsilon^{2}_{k}}\right)\nonumber\\ &-&\mu.\label{eq3}
\end{eqnarray}
It is interesting to see that the Abriksov fermions have the flat band $E_{k\sigma\pm}^{f}$ due to vanished Kondo hybridization $V$. In contrast, the bands of conduction electrons $E_{k\sigma\pm}^{c}$ are dispersive and show the characteristic spin-splitting of AM phase. (Fig.~\ref{fig:2}) In addition, under $C_{4z}$ operation (rotation $\pi/2$ around $z$-axis, $(k_{x},k_{y})\rightarrow(-k_{y},k_{x})$), $E_{k\sigma\pm}^{c}\rightarrow E_{k-\sigma\pm}^{c}$, which implies the $d$-wave symmetry of AM found here. Thus, the spin-resolved Fermi surface in Fig.~\ref{fig:22}(a) also exhibits the $C_{4z}$ symmetry.

The advantage of the simplified Hamiltonian $\hat{H}$ is that one can derive the mean-field equations for $m_{c},m_{f}$ via the following expression of free energy (see Appendix.~\ref{Ap3} for Green's function and spin susceptibility in the pure AM state)
\begin{eqnarray}
f&=&-T\sum_{k\sigma}(\ln(1+e^{-\beta E_{k\sigma+}^{c}})+\ln(1+e^{-\beta E_{k\sigma-}^{c}}))\nonumber\\
&-&T\sum_{k\sigma}(\ln(1+e^{-\beta E_{k\sigma+}^{f}})+\ln(1+e^{-\beta E_{k\sigma-}^{f}}))\nonumber\\
&+&2N_{s}(Jm_{c}m_{f}-\lambda).
\end{eqnarray}
In this formula, $\beta=1/T$ is the inverse temperature.
Now, minimizing free energy $f$ ($\frac{\partial f}{\partial m_{c}}=\frac{\partial f}{\partial m_{f}}=0$) leads to mean-field equations for $m_{c}$ and $m_{f}$:
\begin{eqnarray}
&&m_{f}=\frac{1}{4N_{s}}\sum_{k\sigma}\sigma(f_{F}(E_{k\sigma-}^{f})-f_{F}(E_{k\sigma+}^{f}))\nonumber\\
m_{c}&=&\frac{1}{4N_{s}}\sum_{k\sigma}\sigma\frac{\varepsilon_{k}^{AA}-\varepsilon_{k}^{BB}-Jm_{f}\sigma}{\sqrt{(\varepsilon_{k}^{AA}-\varepsilon_{k}^{BB}-Jm_{f}\sigma)^{2}+4\varepsilon_{k}^{2}}}\nonumber\\
&\times&(f_{F}(E_{k\sigma+}^{c})-f_{F}(E_{k\sigma-}^{c}))\label{MFAM_1}
\end{eqnarray}
Note that $f_{F}(x)=1/(e^{x/T}+1)$ is the standard Fermi distribution function.

The equations for $n_{c}$ and the constraint are also given,
\begin{eqnarray}
&&n_{c}=\frac{1}{2N_{s}}\sum_{k\sigma}(f_{F}(E_{k\sigma+}^{c})+f_{F}(E_{k\sigma-}^{c}))\nonumber\\
&&1=\frac{1}{2N_{s}}\sum_{k\sigma}(f_{F}(E_{k\sigma-}^{f})+f_{F}(E_{k\sigma+}^{f})).\label{MFAM_2}
\end{eqnarray}
For example, if we set $t=1,t_{+}=0.57,t_{-}=0.03,J=0.3,n_{c}=1$ at zero temperature, by solving Eqs.~\ref{MFAM_1} and \ref{MFAM_2}, the resultant quasiparticle bands have been plotted in Fig.~\ref{fig:2}. (the flat bands are not shown) It is amusing to see that in different from the insulating nature of Kondo state at half-filling, the AM state is metallic and satisfies the Luttinger theorem. (see Appendix.~\ref{Ap3} for a brief discussion on Luttinger theorem\cite{Luttinger1960,Oshikawa2000,Dzyaloshinskii2003}) Therefore, the transition from AM to Kondo insulator (if it occurs) behaves as not only a magnetic-paramagnetic transition but also a metal-insulator transition.

Naively, a finite-temperature phase transition out of AM may be obtained if one solves Eqs.~\ref{MFAM_1} and \ref{MFAM_2} for $T>0$. However, the Mermin-Hohenberg theorem prohibits this putative thermodynamic transition since the KLM has the continuous $SU(2)$ spin symmetry. Instead, we expect such finite-$T$ AM transition appears in a three-dimensional KLM or a two-dimensional anisotropic KLM (Ising-Kondo lattice) with alternating NNNH.\cite{WWYang2019b,WWYang2020,WWYang2021}
\subsection{Coexistent state}
After learning the properties of Kondo and AM states, we turn to the case of their coexistence. The coexistent state is expected to be stable in the intermediate regime of $J$ and it is indeed found in our mean-field calculation.

Fig.~\ref{fig:1}(b) shows all three kinds of ground-states and the corresponding order parameters are exhibited in Fig.~\ref{fig:1}(c), particularly, the coexistent phase of Kondo and AM occupies a finite parameter regime.

\begin{figure}
\includegraphics[width=1.0\linewidth]{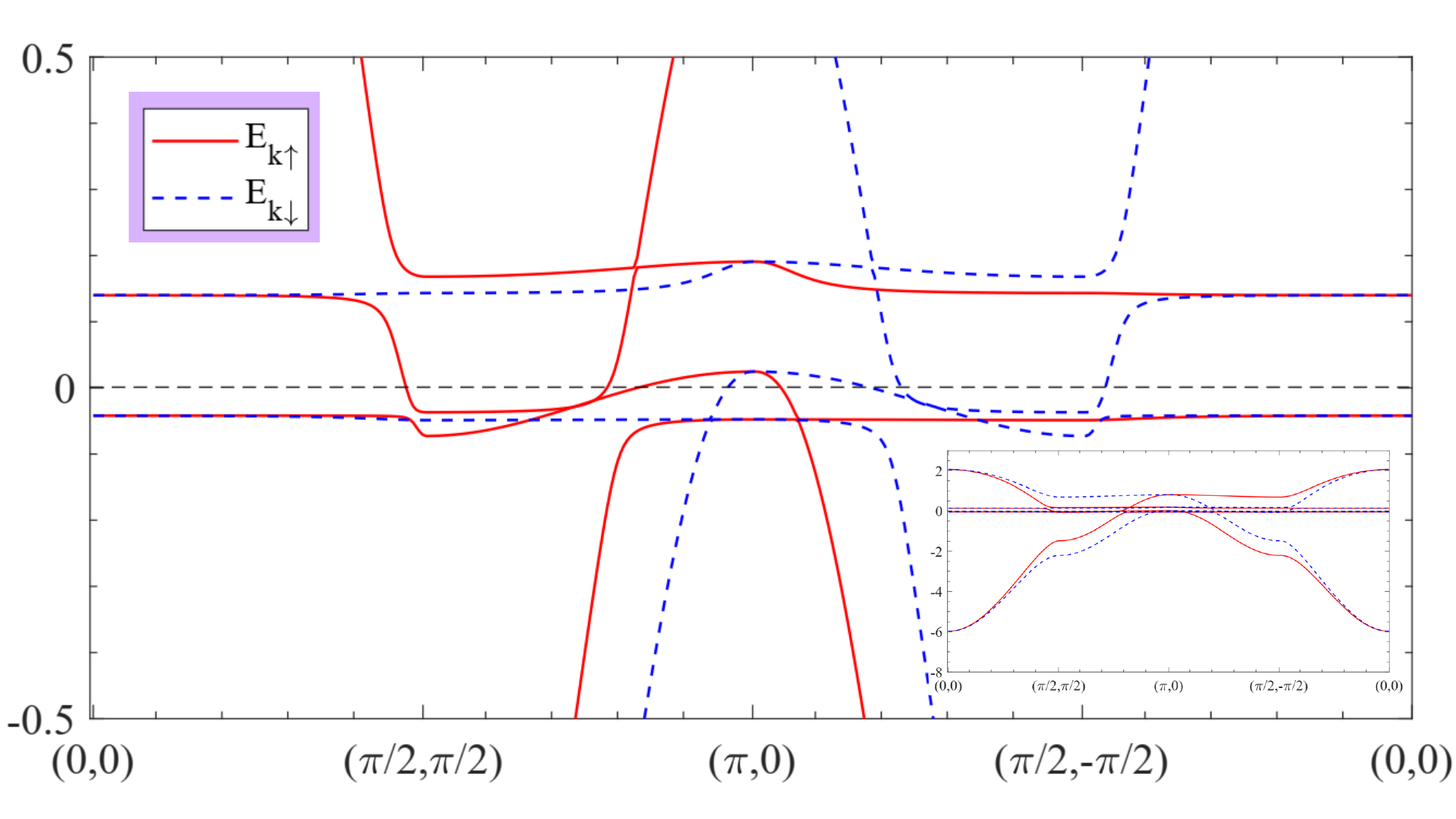}
\caption{\label{fig:4} The spin-splitting bands in the coexistent state of KLM. (The inset figure gives much larger energy range.) The $C_{4z}$ symmetry still works as in AM case. ($t=1,t_{+}=0.57,t_{-}=0.03,J=1.5,n_{c}=1,T=0$)}
\end{figure}

Since $V,m_{c},m_{f}$ are all nonzero in the coexistent phase, we are unable to derive explicit formulas for
quasiparticle bands and mean-field equations, thus certain numerical iterative algorithm has to be used. A case study has been shown in Fig.~\ref{fig:4}, where all of the eight quasiparticle bands $E_{k\uparrow},E_{k\downarrow}$ are plotted. As expected, these bands are spin-splitting as a result of non-vanishing contribution of AM order. At the same time, they are all dispersive and have no flat bands. This fact is due to the hybridization of Kondo screening in the coexistent state. The hybridization mixes the original flat bands in AM with dispersive bands of conduction electrons. Additionally, the $C_{4z}$ symmetry is preserved as in the AM case, so the coexistent state also realizes a $d$-wave AM.

\begin{figure*}
\includegraphics[width=0.65\linewidth]{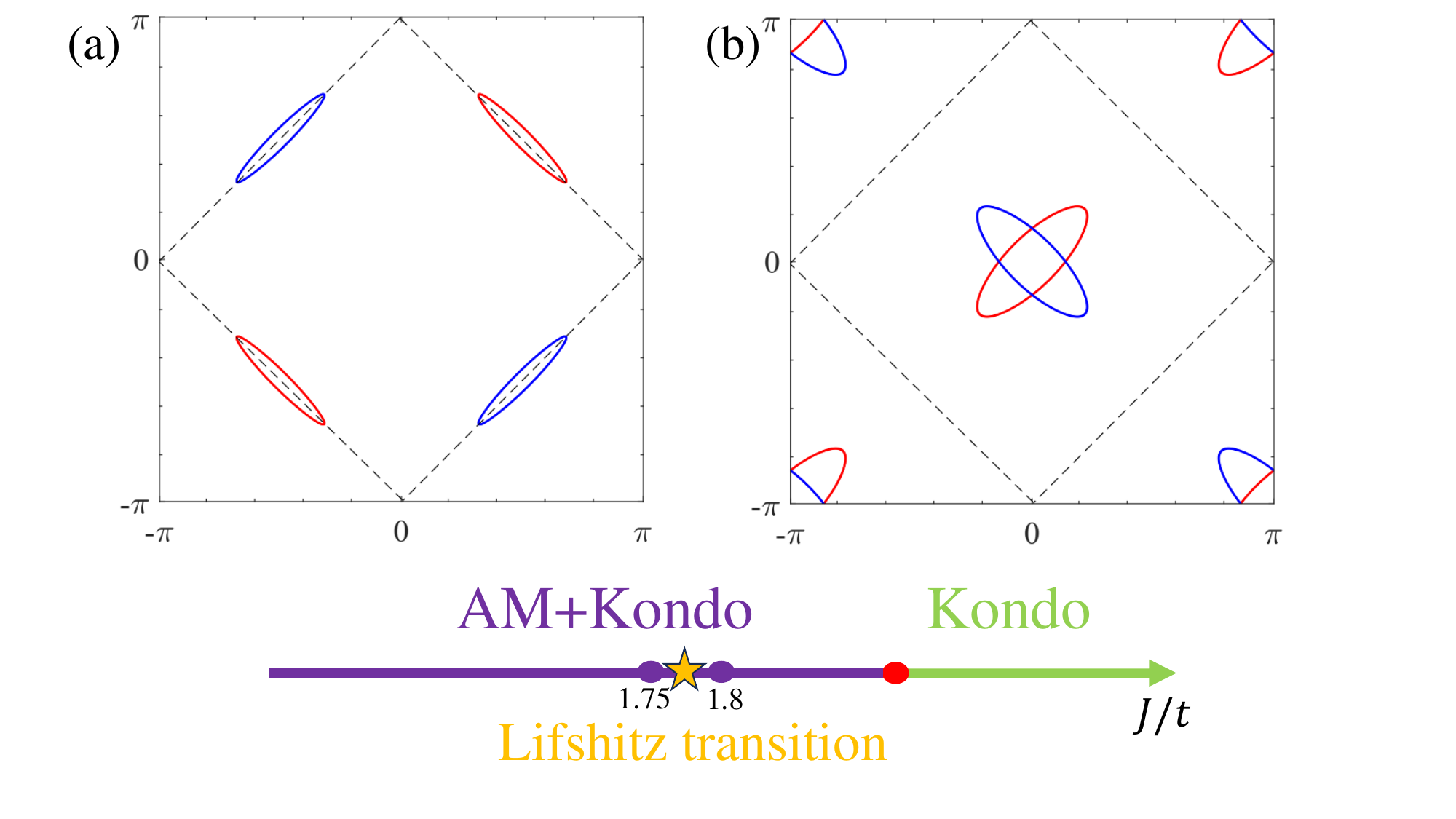}
\caption{\label{fig:6} The Fermi surface of quasiparticle in the coexistent state. (a)$J/t=1.75$ versus (b) $J/t=1.8$. ($t=1,t_{+}=0.57,t_{-}=0.03,n_{c}=1,T=0$)}
\end{figure*}

An interesting feature from Fig.~\ref{fig:1}(c) is that the order parameters have a jump behavior around $J/t=1.76$. By examining the quasiparticle bands and the Fermi surface, we find such jump is a result of Lifshitz transition, where the topology of Fermi surface radically changes but without any symmetry-breaking.\cite{Continentino} (Fig.~\ref{fig:6}) (Fermi surface centred around $(\pm\pi/2,\pm\pi/2)$ transits into the one centred at $(0,0)$) This Lifshitz transition is different from the short-ranged antiferromagnetic correlation-driven counterpart in Kondo-Heisenberg model.\cite{GMZhang2011} Interestingly, the spin-splitting Fermi surface (Fig.~\ref{fig:6}(b)) after the Lifshitz transition has the similar structure of $d_{xy}$-wave ($\sim\sigma\sin k_{x}\sin k_{y}$), which is relevant to the AM material RuO$_{2}$.\cite{Smejkal2022}

After all, we have uncovered $d$-wave AM-like phases in the half-filled KLM. The defining feature of AM, the spin-splitting bands are unambiguously observed in our mean-field calculation. We emphasize that these states are driven by the antiferromagnetic magnetic order of local spin moments. If local spin moments lose the antiferromagnetic order (due to magnetic/geometric frustration), no AM-like states will survive. For example, when local spin moments form valence-bond solids or quantum spin liquids,\cite{Zhou2017} the antiferromagnetic long-ranged order is totally lost and the system may be in the fractionalized Fermi liquid.\cite{Senthil2003}

\subsection{Phase transitions in phase diagram}
We have clarified basic features of three candidate ground-states. From Fig.~\ref{fig:1}, there are transitions between AM and coexistent state, coexistent state and Kondo state.

The phase transition between AM and coexistent state is characterized by the appearance of Kondo screening ($V\neq0$). The corresponding effective field theory may be the one established in the study of Kondo-breakdown mechanism although the effect of the magnetic background is not clear.\cite{Senthil2004,Paul2007,Vojta2010,Zhong2012} (Frankly speaking, a small dispersion of Abriksov fermions has to be added if one considers this seriously.) As discussed in Appendix.~\ref{Ap4}, the corresponding effective action is
\begin{equation}
S_{eff}=\sum_{\Omega_{n},q}\left(J-(J/2)^{2}\chi_{0}^{\alpha}(q,\Omega_{n})\right)|V_{q\alpha}(\Omega_{n})|^{2}
\end{equation}
where $V_{q\alpha}(\Omega_{n})$ denotes the Kondo order field. In the sense of Landau symmetry-breaking theory, we may treat Kondo order field as the order parameter and its condensation leads to the coexistent state, where the Fermi surface is enlarged by hybridizing with Abriksov fermions.\cite{Senthil2004,Paul2007,Vojta2010,Zhong2012} In contrast, if $V$ does not condense, the system is in the AM state and the Abriksov fermions have no contribution to the Fermi surface.
In light of above features, AM may be called a local moment magnet while the coexistent state is an itinerant magnet. This fact underlies that $f$-electron is localized in the former case and it hybridizes with conduction electron in the latter one.

Finally, the coexistent state transits into the Kondo state when $J$ is large. In this situation, magnetic order parameters $m_{c},m_{f}$ gradually decrease and vanish at the second-order transition point. When magnetic fluctuations are included, it is expected that this transition can be described by the classic Hertz-Millis theory with the dynamic critical exponent $z=2$.\cite{Hertz,Millis} Here, $z=2$ is related to the critical Landau-damped paramagnetic excitation around Fermi surface with characteristic wavevector $Q=(\pi,\pi)$.

Before ending this subsection, we note that the pure AM state, predicted by the present mean-field theory, may disappear if one considers its counterpart in KLM without NNNH on square lattice.\cite{Raczkowski2022} If such AM state indeed exists, there will be a Kondo-breakdown transition between AM and the coexistent state.\cite{Senthil2004,Paul2007,Vojta2010,Zhong2012} This non-Landau symmetry-breaking phase transition is more possibly realized on honeycomb lattice,\cite{Zhong2013,YLiu2017} where low-lying Dirac fermions are susceptible to fluctuations neglected in the mean-field approximation.

\subsection{Some observable}
Except the spin-splitting bands and Fermi surface, let us inspect other physical observable for the states involving AM orders.

\subsubsection{Spin-resolved density of states}
The spin-resolved density of states is defined by
\begin{eqnarray}
N_{\sigma}(\omega)=-\frac{1}{\pi N_{s}}\sum_{k}\mathrm{Tr}\mathrm{Im}G_{\sigma}^{R}(k,\omega)\nonumber\\
\end{eqnarray}
and the retarded Green's function is defined as
\begin{eqnarray}
G_{\sigma}^{R}(k,\omega)=(\omega+i0^{+}-H_{\sigma}(k))^{-1},
\end{eqnarray}
which can be detected by the state-of-art spin-resolved scanning tunnel microscope (STM).\cite{Wiesendanger2009,Bode2003} However, due to the intrinsic $C_{4z}$ symmetry in $d$-wave AM-like states, one finds $N_{\uparrow}(\omega)=N_{\downarrow}(\omega)$. Thus, it seems that the spin-resolved density of states is not able to identify AM-like phases.
\subsubsection{Spin-resolved conductivity}
The spin-splitting bands are responsible for the spin-resolved conductivity in AM-like phases, which is essential to spintronics, such as the spin splitting torque effect.\cite{Smejkal2022,Bai2022,Feng2024} According to the linear-response theory, the zero-temperature conductivity has the following expression,
\begin{equation}
\sigma_{\alpha\alpha}^{\sigma}=\frac{e^{2}}{\hbar}\frac{1}{N_{s}}\sum_{k}\mathrm{Tr}\left[v_{\alpha}^{\sigma}(k)
A_{\sigma}(k)v_{\alpha}^{\sigma}(k)A_{\sigma}(k)\right].
\end{equation}
Here, the velocity along $(1,1)$-direction is $v_{\tilde{x}}^{\sigma}(k)=\frac{1}{\sqrt{2}}(\partial_{k_{x}}+\partial_{k_{y}})H_{\sigma}(k)$ while $v_{\tilde{y}}^{\sigma}(k)=\frac{1}{\sqrt{2}}(\partial_{k_{x}}-\partial_{k_{y}})H_{\sigma}(k)$ is the velocity along $(1,-1)$-direction. The zero-frequency spectral function $A_{\sigma}(k)=-\frac{1}{\pi}\mathrm{Im}G_{\sigma}^{R}(k,\omega=0)$. Due to the $C_{4z}$ symmetry of bands in AM-like phases, one expects $\sigma_{\tilde{x}\tilde{x}}^{\uparrow}=\sigma_{\tilde{y}\tilde{y}}^{\downarrow},\sigma_{\tilde{x}\tilde{x}}^{\downarrow}=\sigma_{\tilde{y}\tilde{y}}^{\uparrow}$.
(The conductivity along $(1,0)$ and $(0,1)$-direction obeys $\sigma_{xx}^{\uparrow}=\sigma_{xx}^{\downarrow}=\sigma_{yy}^{\uparrow}=\sigma_{yy}^{\downarrow}$ because of $C_{4z}$-rotation and mirror operation) More specifically, in the AM state with $J/t=0.3$, we find $\sigma_{\tilde{x}\tilde{x}}^{\uparrow}=\sigma_{\tilde{y}\tilde{y}}^{\downarrow}=2.605\frac{e^{2}}{\hbar}$, $\sigma_{\tilde{x}\tilde{x}}^{\downarrow}=\sigma_{\tilde{y}\tilde{y}}^{\uparrow}=0.445\frac{e^{2}}{\hbar}$. The coexistent state ($J/t=1.5$) gives $\sigma_{\tilde{x}\tilde{x}}^{\uparrow}=\sigma_{\tilde{y}\tilde{y}}^{\downarrow}=3.877\frac{e^{2}}{\hbar}$, $\sigma_{\tilde{x}\tilde{x}}^{\downarrow}=\sigma_{\tilde{y}\tilde{y}}^{\uparrow}=0.099\frac{e^{2}}{\hbar}$. (a damping factor $\Gamma=0.01t$ is used to obtain finite conductivity and the system size is $300\times300$) $\sigma_{\alpha\alpha}^{\uparrow}\neq\sigma_{\alpha\alpha}^{\downarrow}$ implies a nonzero spin polarized current, thus heavy fermion compounds with AM-like states have the potential for spintronics applications.
\subsubsection{Distribution of conduction electron}
\begin{figure}
\includegraphics[width=1.0\linewidth]{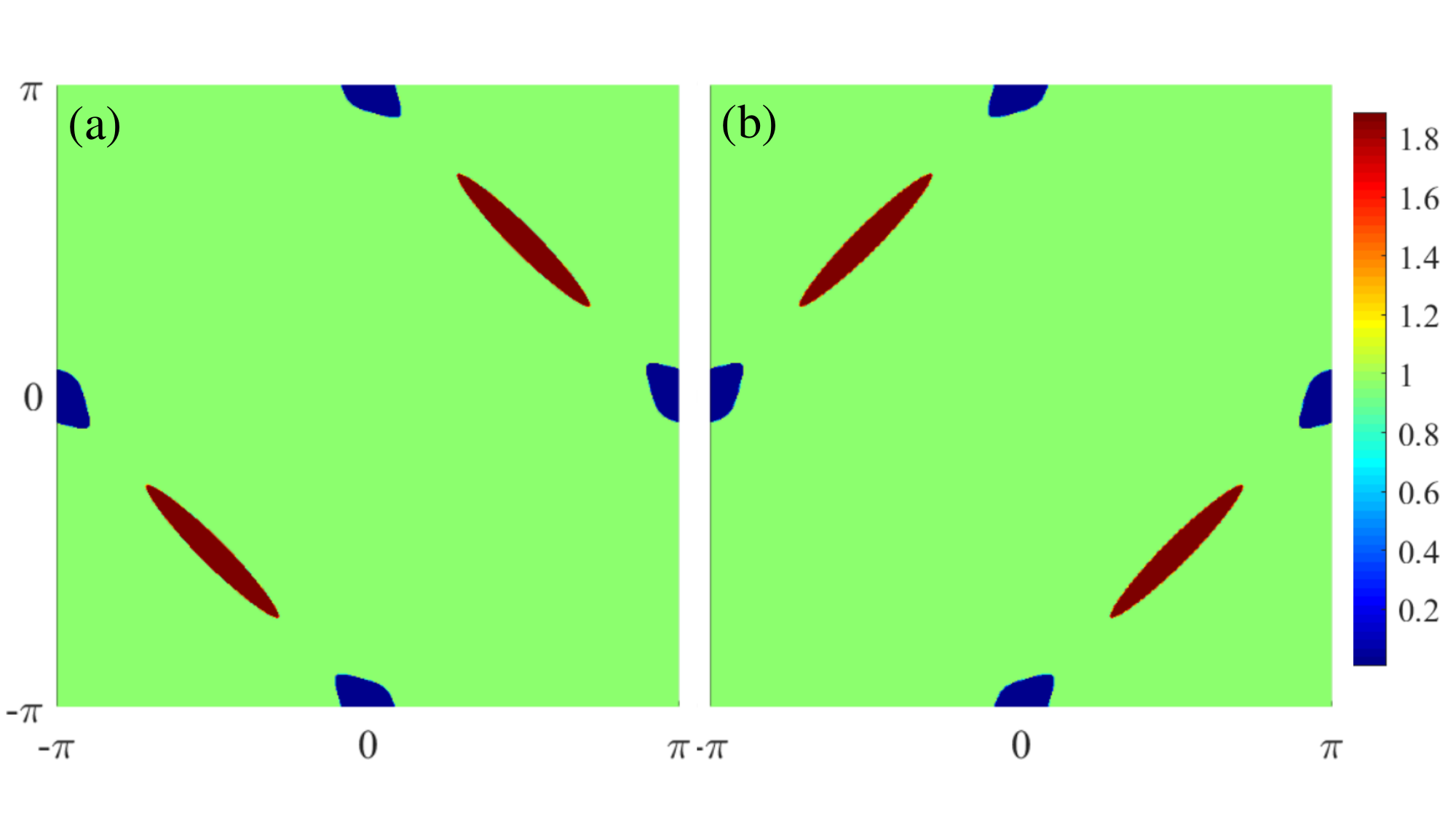}
\caption{\label{fig:11} The distribution function of conduction electron (a) $n_{k\uparrow}^{c}$ and (b) $n_{k\downarrow}^{c}$ in the AM state.($t=1,t_{+}=0.57,t_{-}=0.03,J=0.3,n_{c}=1,T=0$)}
\end{figure}

\begin{figure}
\includegraphics[width=1.0\linewidth]{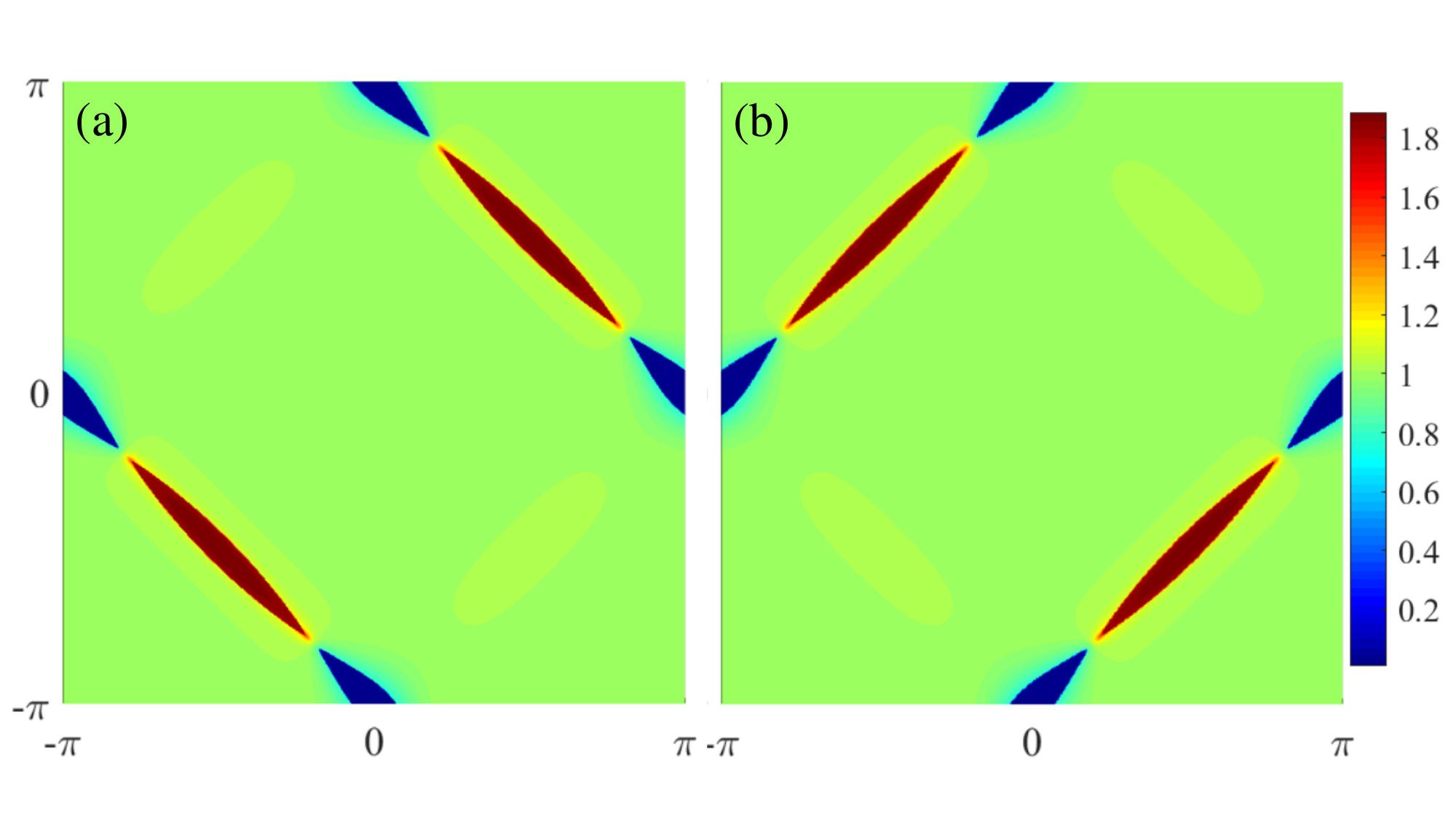}
\caption{\label{fig:12} The distribution function of conduction electron (a) $n_{k\uparrow}^{c}$ and (b) $n_{k\downarrow}^{c}$ in the coexistent  state.($t=1,t_{+}=0.57,t_{-}=0.03,J=1.5,n_{c}=1,T=0$)}
\end{figure}

The distribution function of conduction electron $n_{k\sigma}^{c}=\langle \hat{c}_{kA\sigma}^{\dag}\hat{c}_{kA\sigma}\rangle+\langle \hat{c}_{kB\sigma}^{\dag}\hat{c}_{kB\sigma}\rangle$ is shown in Figs.~\ref{fig:11} and \ref{fig:12}. It is evident that $n_{k\sigma}^{c}$ is spin-resolved in AM-like phases and gives the same information just as the spin-splitting Fermi surface. The advantage of distribution function over the spin-splitting Fermi surface/bands is that encoding the latter requires calculating Green's function of conduction electron, which is rather demanding for many-body techniques due to its frequency-dependence. In contrast, calculating $n_{k\sigma}^{c}$ is much easier.

\section{Discussion}\label{sec3}
\subsection{Implication for experiment and realistic materials: how to detect AM phases in heavy fermions}
The key feature of AM-like phases in KLM or generic heavy fermion systems is obviously the spin-splitting bands. These bands can be observed by ARPES as in the case of $d$-electron compounds RuO$_{2}$ and MnTe.\cite{Krempasky2024,Fedchenko2024,Lee2024,Osumi2024}

Because the AM-like states are driven by antiferromagnetic magnetic order of local spin moments, exploring AM in heavy fermions can be considered as a task to search antiferromagnetic heavy fermion compounds. However, the magnetic transition temperature of heavy fermion materials is rather low (usually several Kelvin),\cite{Kaczorowski2010,YLuo2011} which is a challenge for current ARPES techniques.

In contrast, the low-temperature magnetic quantum oscillation widely used in heavy fermion materials is able to probe the Landau quantization of quasiparticles in applied magnetic field induced by the spin-splitting Fermi surface, related to the spin-splitting bands.\cite{Shoenberg,Tan2015,Maksimovic2022} Thus, AM-like phases are more likely to be detected by the quantum oscillation measurements, performed on thermodynamics and/or transport.\cite{ZXLi2024,YHuang2024}

We have seen that the spin-resolved density of states detected by STM is not able to identify AM-like phases. However, if one focus on the spatial-dependent density of states around impurity, existing in any real materials, the configuration of those density of states can provide information on the underlying Fermi surface, thus it acts as a tool for exploring AM-like phases.\cite{WChen2024,Sukhachov2024}

Additionally, the low-temperature transport measurement in heavy fermion materials is standard, so AM-like phases could be detected by charge/spin/thermal transport just like the spin-resolved conductivity calculated in last section.
\subsection{Alternative theoretical approach}
\subsubsection{Some mean-field theories}
In the main text, we have used fermionic parton mean-field theory to explore AM-like phase in KLM on square lattice. Other mean-field theories like bond-operator and Schwinger boson may also work.\cite{Jurecka2001,Eder2018,Komijani2018,JWang2021,RHan2021} There seems to be no pure antiferromagnetic phase in the bond-operator approach, so we expect AM to coexist with Kondo screening in this situation.

The (dynamic) large-N theory of Schwinger boson has the same spirit with the more conventional non-crossing approximation, which is widely used in the study of single impurity Anderson/Kondo model.\cite{Hewson1993} It is not clear to us whether it will work well for the coexistent state with both AM and Kondo screening.
\subsubsection{Quantum Monte Carlo simulation}
Theoretical tools beyond mean-field approximation are required if one desires to have more exact information about possible AM and related phases. The numerically exact determinant quantum Monte Carlo simulation works only for half-filled KLM without NNNH.\cite{Assaad1999} When NNNH is nonzero, particle-hole symmetry breaks and the corresponding fermion sign-problem prevents reliable calculations for large lattice size or low-temperature. In this direction, we note that the constrained path Monte Carlo (CPMC) method may be applicable if one starts with a well-tested trial wavefunction.\cite{SZhang1995,MQin2020}

A promising way to search AM-like phase is to study the following periodic Anderson model with alternating NNNH,
\begin{eqnarray}
\hat{H}&=&-t\sum_{i,\delta,\sigma}(\hat{c}_{iA\sigma}^{\dag}\hat{c}_{i+\delta,B\sigma}+\hat{c}_{i+\delta,B\sigma}^{\dag}\hat{c}_{iA\sigma})\nonumber\\
&-&\sum_{i,\delta_{1}',\sigma}(t_{+}\hat{c}_{iA\sigma}^{\dag}\hat{c}_{i+\delta_{1}',A\sigma}+t_{-}\hat{c}_{iB\sigma}^{\dag}\hat{c}_{i+\delta_{1}',B\sigma})\nonumber\\
&-&\sum_{i,\delta_{2}',\sigma}(t_{-}\hat{c}_{iA\sigma}^{\dag}\hat{c}_{i+\delta_{2}',A\sigma}+t_{+}\hat{c}_{iB\sigma}^{\dag}\hat{c}_{i+\delta_{2}',B\sigma})\nonumber\\
&+&E_{f}\sum_{i,\sigma}(\hat{f}_{iA\sigma}^{\dag}\hat{f}_{iA\sigma}+\hat{f}_{iB\sigma}^{\dag}\hat{f}_{iB\sigma})\nonumber\\
&+&U\sum_{i}(\hat{f}_{iA\uparrow}^{\dag}\hat{f}_{iA\uparrow}\hat{f}_{iA\downarrow}^{\dag}\hat{f}_{iA\downarrow}+\hat{f}_{iB\uparrow}^{\dag}\hat{f}_{iB\uparrow}\hat{f}_{iB\downarrow}^{\dag}\hat{f}_{iB\downarrow})\nonumber\\
&+&V\sum_{i,\sigma}(\hat{f}_{iA\sigma}^{\dag}\hat{c}_{iA\sigma}+\hat{f}_{iB\sigma}^{\dag}\hat{c}_{iB\sigma}+h.c.).
\end{eqnarray}
Here, $\hat{f}_{iA\sigma},\hat{f}_{iB\sigma}$ represent the local $f$-electron orbit. The $U$ term is the familiar Hubbard interaction for on-site $f$-electrons and $E_{f}$ is the energy level of $f$-electron. A hybridization $V$ between conduction electron and $f$-electron is crucial since it leads to Kondo screening and other many-body effect. In the so-called Kondo limit ($E_{f}\ll E_{F}\ll E_{f}+U, V\ll U$ with $E_{F}$ being Fermi energy), the above model can be mapped into KLM we have studied in the main text. We think it is highly desirable to search the physics of AM in above model in terms of the CPMC method.

\section{Conclusion and Future direction}\label{sec4}
In conclusion, we have uncovered AM-like phases in a prototypical model of heavy fermion, the KLM on square lattice with an alternating NNNH. The introduction of alternating NNNH takes nonmagnetic atoms (neglected in usual antiferromagnetism study) into account and it underlies the $\pi/2$-rotation in addition to the well-understood time-reversal and translation operation in usual antiferromagnetism. The $d$-wave AM-like magnetic states are characterized by their spin-splitting bands, Fermi surface, spin-resolved distribution function and conductivity. Inspired by AM materials such as RuO$_{2}$ with rotation-connected sublattices, we believe searching AM states in heavy fermion compounds is promising when low-temperature transport and magnetic quantum oscillation measurements are performed on antiferromagnetic heavy fermion materials with rotation-connected sublattices.

When Heisenberg interaction between local moments is added, KLM is generalized to the Kondo-Heisenberg model. Extension of our mean-field theory should predict the existence of superconductivity due to the resonance-valence-bond mechanism and its interplay with AM order will lead to topological superconductivity.\cite{Zhu2023} This interesting issue will be explored in our future work. Although AM phases found in the present work root on a specific mean-field calculation, we hope they could be discovered in real-life $f$-electron compounds.

\emph{Note added}: After submitting the present manuscript, we have noted two groups' LDA calculations, which predict that $f$-electron CeNiAsO and Ce$_{4}$X$_{3}$ (X=As,Sb,Bi) are candidate of AM.\cite{Hellenes2024,XHe2024} Particularly, the former one gives an example of $p$-wave AM though it requires a coplanar magnetic configuration. We expect that the present mean-field theory with suitable extension may capture the key feature of $p$-wave AM state in CeNiAsO.
\section*{Acknowledgments}
We thank Yu Li for bringing the author's attention to the topic of altermagnet. The work is partly supported by
the National Key Research and Development Program of China (Grant No. 2022YFA1402704) and the programs for NSFC of China (Grant No. 11834005, Grant No. 12247101). This research was supported in part by Supercomputing Center of Lanzhou University. We thank the Supercomputing
Center of Lanzhou University for allocation of CPU time.

\appendix

\section{Examples with other $t_{+},t_{-}$}\label{Ap1}
\subsection{$t_{+}=t_{-}=0$}
\begin{figure}
\includegraphics[width=1.0\linewidth]{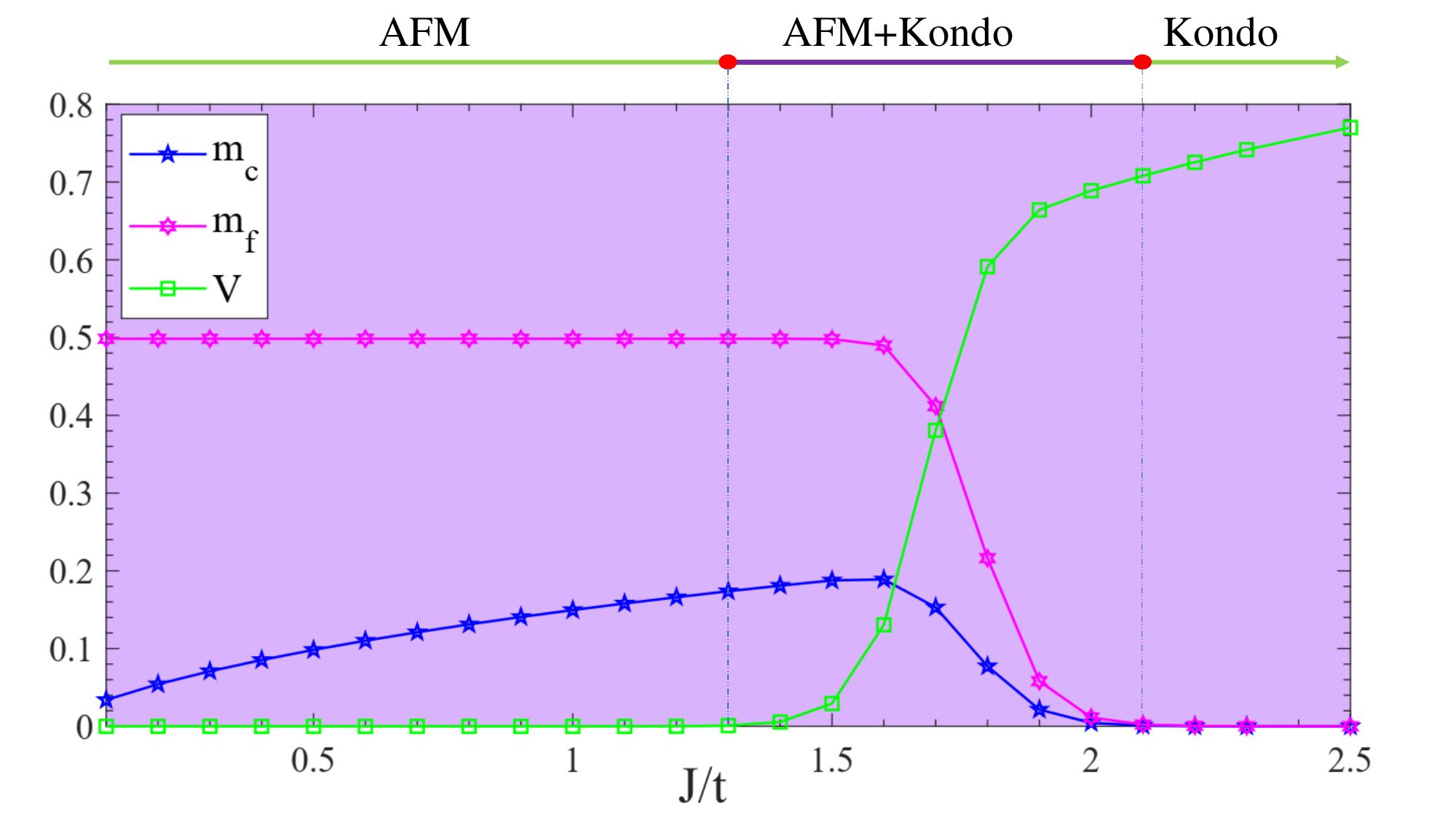}
\caption{\label{fig:7} Order parameters and phase diagram in half-filled KLM without NNNH. ($t=1,t_{+}=t_{-}=0,n_{c}=1,T=0$)}
\end{figure}

If the NNNH is turned off, ($t_{+}=t_{-}=0$) the model is just the symmetric KLM on square lattice. The mean-field phase diagram and order parameters at half-filling are shown in Fig.~\ref{fig:7}. Here, AFM denotes the conventional antiferromagnetic state and it can coexist with the Kondo screening state. The critical point between the coexistent state and the Kondo state is around $J/t\simeq2.1$, which is not too different from the exact value $J/t=1.45\pm0.05$ extracted from the quantum Monte Carlo simulation.\cite{Assaad1999} (Note that a different large-N mean-field theory used in Ref.~\onlinecite{Li2015} gives a much larger value $J/t\simeq3.4$.) In this case, we have checked that all quasiparticle bands in all three phases are not spin-splitting and no AM-like phases appear as expected.

\subsection{$t_{+}=0.4,t_{-}=0.2$}
\begin{figure}
\includegraphics[width=1.0\linewidth]{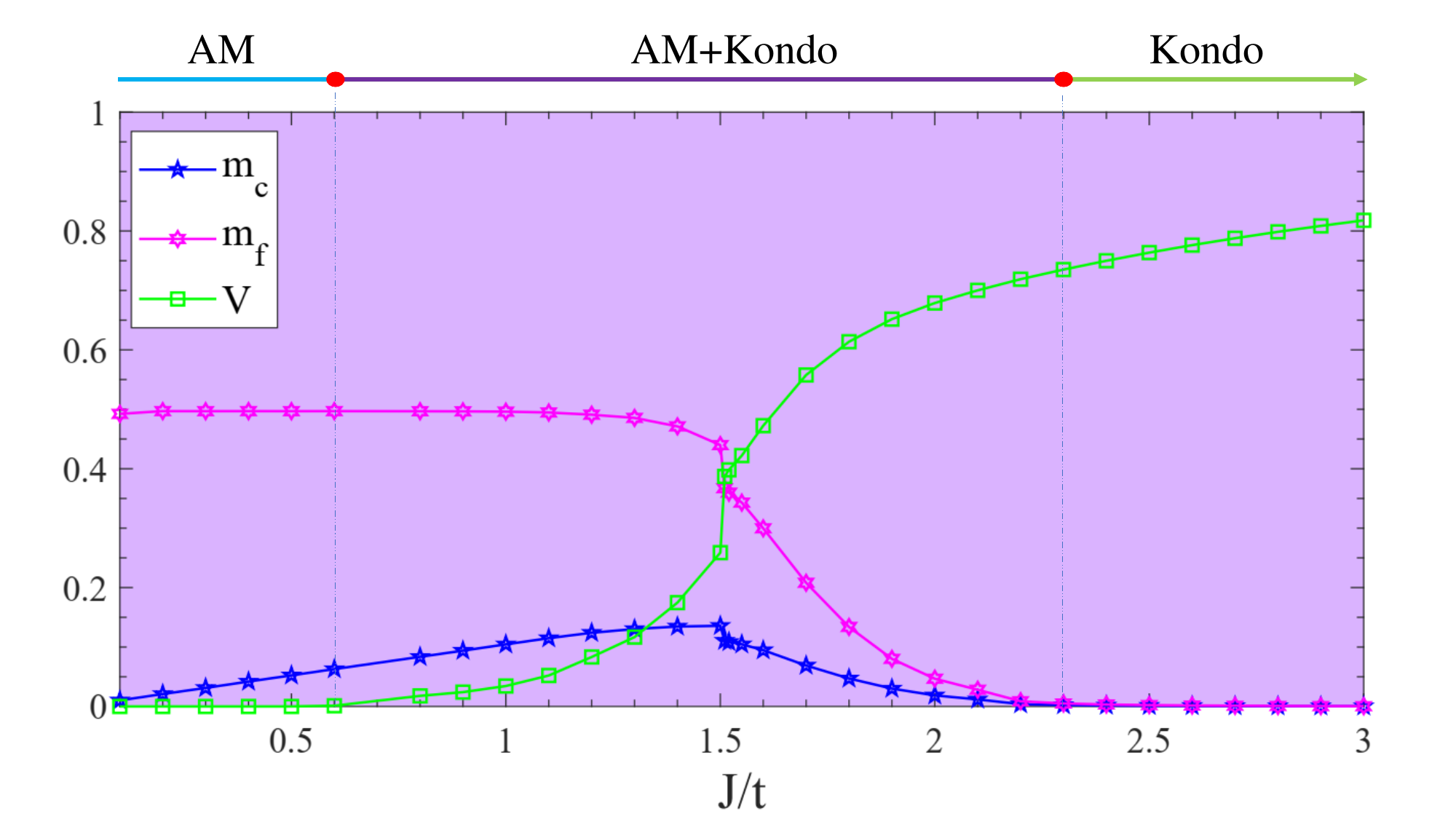}
\caption{\label{fig:8} Order parameters and phase diagram in half-filled KLM with NNNH. ($t=1,t_{+}=0.4,t_{-}=0.2,n_{c}=1,T=0$)}
\end{figure}

\begin{figure}
\includegraphics[width=1.0\linewidth]{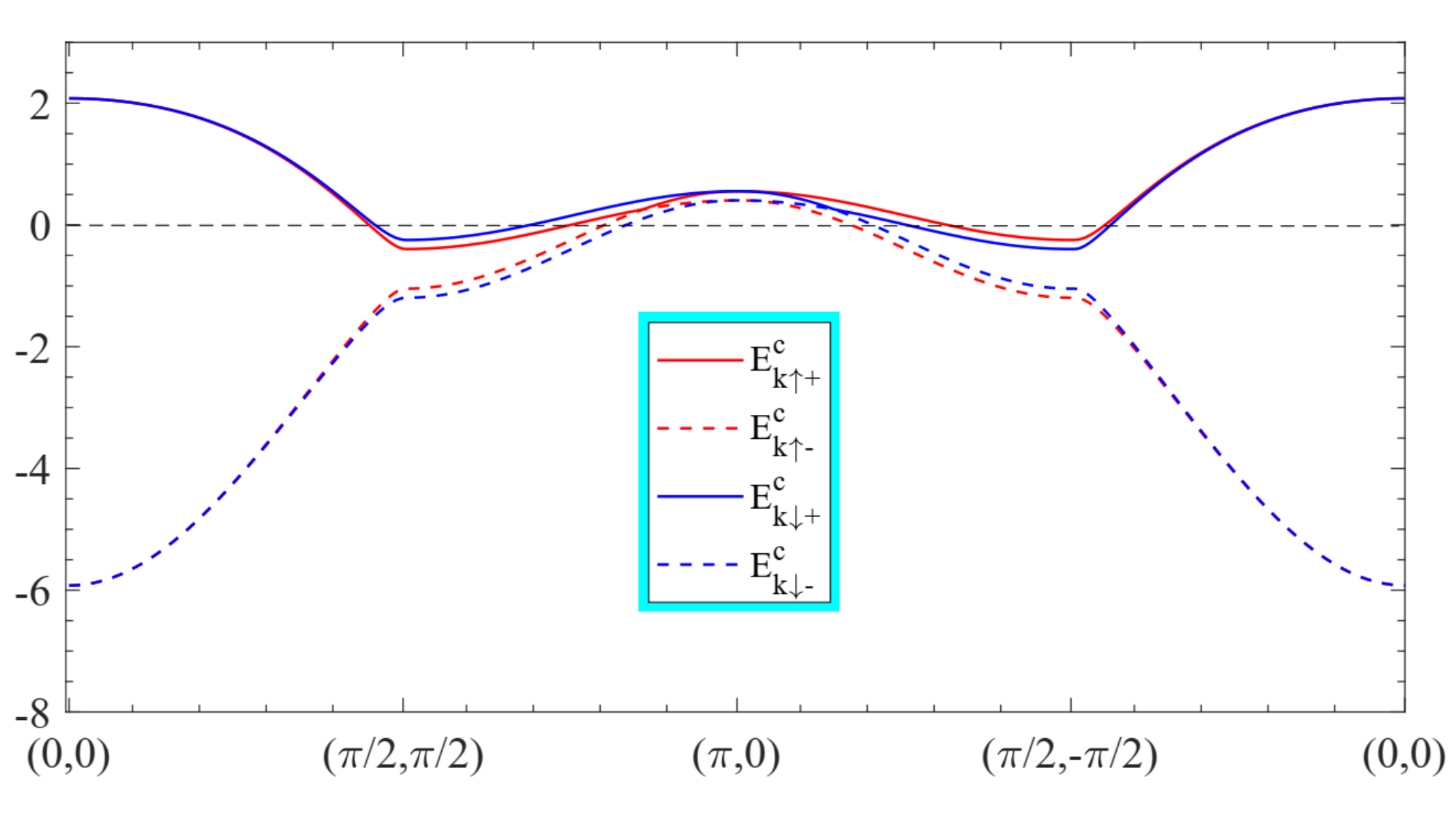}
\caption{\label{fig:9} The spin-splitting bands in the AM state of KLM. The $C_{4z}$ symmetry of bands implies a $d$-wave AM. ($t=1,t_{+}=0.4,t_{-}=0.2,J=0.3,n_{c}=1,T=0$)}
\end{figure}

\begin{figure}
\includegraphics[width=1.0\linewidth]{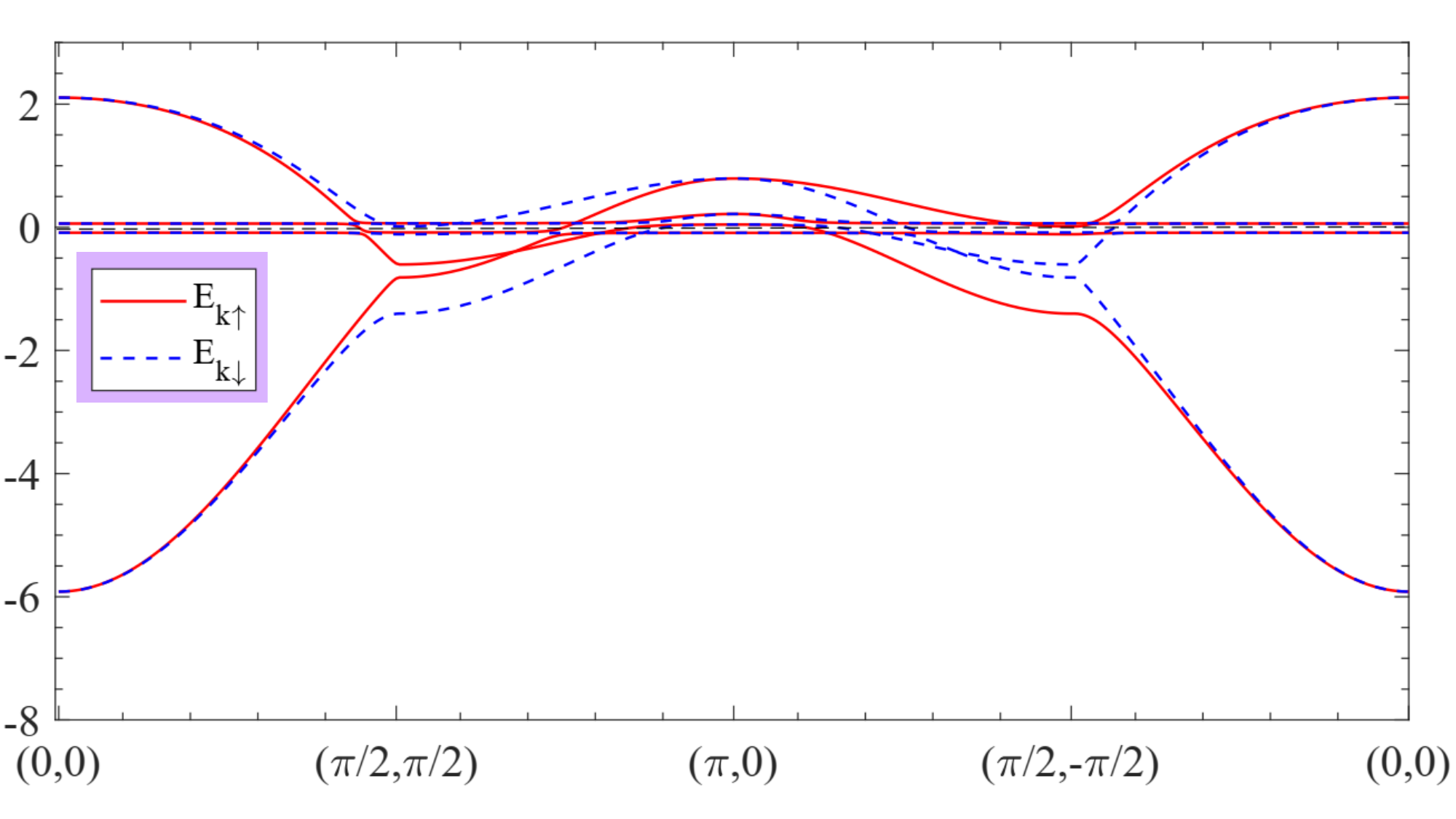}
\caption{\label{fig:10} The spin-splitting bands in the coexistent state of KLM. The $C_{4z}$ symmetry of bands implies a $d$-wave AM. ($t=1,t_{+}=0.4,t_{-}=0.2,J=1.2,n_{c}=1,T=0$)}
\end{figure}

Then, let us consider NNNH is not zero and we set $t_{+}=0.4,t_{-}=0.2$. In Fig.~\ref{fig:8}, we see that there exist three kinds of states just as the findings in the main text. The AM and the coexistent states are also $d$-wave-like, as what can be seen from Figs.~\ref{fig:9} and \ref{fig:10}. Additionally, a Lifshitz transition appears around $J/t=1.51$, which is also found in the main text. Therefore, we believe the results in the main text is robust for generic parameters.
\section{Green's function and spin susceptibility in the pure AM state}\label{Ap3}
Here, we provide conduction electron Green's function and spin susceptibility in the pure AM state, which is based on the mean-field theory used in the main text.

In the pure AM state without Kondo screening, with the help of mean-field Hamiltonian Eq.~\ref{eq2}, the Green's function of conduction electron on the $A$-sublattice is easy to find as
\begin{eqnarray}
G_{A\sigma}^{c}(k,\omega_{n})&=&\frac{i\omega_{n}-(\varepsilon_{k}^{BB}-\mu+J\sigma S/2)}{(i\omega_{n}-E_{k\sigma+})(i\omega_{n}-E_{k\sigma-})}\nonumber\\
&=&\frac{\mu_{k\sigma}^{2}}{i\omega_{n}-E_{k\sigma+}}+\frac{\nu_{k\sigma}^{2}}{i\omega_{n}-E_{k\sigma-}},
\end{eqnarray}
while the Green's function on the $B$-sublattice is
\begin{equation}
G_{B\sigma}^{c}(k,\omega_{n})=\frac{\nu_{k\sigma}^{2}}{i\omega_{n}-E_{k\sigma+}}+\frac{\mu_{k\sigma}^{2}}{i\omega_{n}-E_{k\sigma-}}.
\end{equation}
Here, the coherent factors are defined as
\begin{equation}
\mu_{k\sigma}^{2}=\frac{E_{k\sigma+}-(\varepsilon_{k}^{BB}-\mu+J\sigma S/2)}{E_{k\sigma+}-E_{k\sigma-}},
\end{equation}
\begin{equation}
\nu_{k\sigma}^{2}=\frac{(\varepsilon_{k}^{BB}-\mu+J\sigma S/2)-E_{k\sigma-}}{E_{k\sigma+}-E_{k\sigma-}},
\end{equation}
and they satisfy $\mu_{k\sigma}^{2}+\nu_{k\sigma}^{2}=1$. The poles of $G_{A\sigma}^{c}(k,\omega_{n}),G_{b\sigma}^{c}(k,\omega_{n})$ determine the quasiparticle bands. The zero-frequency part of these Green's function contributes to the Luttinger theorem\cite{Luttinger1960,Oshikawa2000,Dzyaloshinskii2003}
\begin{equation}
n_{c}=\frac{1}{N_{s}}\sum_{k\sigma}\theta(G_{A\sigma}^{c}(k,0))=\frac{1}{N_{s}}\sum_{k\sigma}\theta(G_{B\sigma}^{c}(k,0)),
\end{equation}
where $\theta(x)$ is the unit-step function. ($\theta(x)=1$ for $x>0$, others it gives zero) We have checked that the Luttinger theorem is satisfied in the pure AM phase.

The (dynamic) spin susceptibility of conduction electron on $A$-sublattice is
\begin{equation}
\chi_{A}^{-+}(q,\Omega_{n})=-\frac{1}{N_{s}}\sum_{k}\frac{1}{\beta}\sum_{\omega_{n}}G_{A\uparrow}^{c}(k,\omega_{n})G_{A\downarrow}^{c}(k+q,\omega_{n}+\omega_{n}),
\end{equation}
and its explicit expression is the following one (using $G_{A\sigma}^{c}(k,\omega_{n})$ and summing over frequency $\omega_{n}$)
\begin{eqnarray}
\chi_{A}^{-+}(q,\Omega_{n})&=&-\frac{1}{N_{s}}\sum_{k}(\mu_{k\uparrow}^{2}\mu_{k+q\downarrow}^{2}\frac{f_{F}(E_{k\uparrow+})-f_{F}(E_{k+q\downarrow+})}{i\Omega_{n}-E_{k+q\downarrow+}+E_{k\uparrow+}}\nonumber\\
&+&\nu_{k\uparrow}^{2}\nu_{k+q\downarrow}^{2}\frac{f_{F}(E_{k\uparrow-})-f_{F}(E_{k+q\downarrow-})}{i\Omega_{n}-E_{k+q\downarrow-}+E_{k\uparrow-}}\nonumber\\
&+&\mu_{k\uparrow}^{2}\nu_{k+q\downarrow}^{2}\frac{f_{F}(E_{k\uparrow+})-f_{F}(E_{k+q\downarrow-})}{i\Omega_{n}-E_{k+q\downarrow-}+E_{k\uparrow+}}\nonumber\\
&+&\nu_{k\uparrow}^{2}\mu_{k+q\downarrow}^{2}\frac{f_{F}(E_{k\uparrow-})-f_{F}(E_{k+q\downarrow+})}{i\Omega_{n}-E_{k+q\downarrow+}+E_{k\uparrow-}}).
\end{eqnarray}
It is seen that all these terms correspond to inter-band transition.

\section{Field theory and effective action}\label{Ap4}

Based on fermionic Hamiltonian Eqs.~\ref{eq4} and \ref{eq5}, one can obtain the following imaginary-time action in terms of coherent states of fermions
\begin{widetext}
\begin{eqnarray}
S&=&\int_{0}^{\beta}d\tau \sum_{j,\alpha=A,B,\sigma}\left(\bar{c}_{j\alpha\sigma}(\partial_{\tau}-\mu)c_{j\alpha\sigma}+\bar{f}_{j\alpha\sigma}(\partial_{\tau}+i\lambda_{j\alpha})f_{j\alpha\sigma}\right)-t\sum_{j,\delta,\sigma}(\bar{c}_{jA\sigma}c_{j+\delta,B\sigma}+\bar{c}_{j+\delta,B\sigma}c_{jA\sigma})\nonumber\\
&-&\sum_{j,\delta_{1}',\sigma}(t_{+}\bar{c}_{jA\sigma}c_{j+\delta_{1}',A\sigma}+t_{-}\bar{c}_{jB\sigma}c_{j+\delta_{1}',B\sigma})-\sum_{j,\delta_{2}',\sigma}(t_{-}\bar{c}_{jA\sigma}c_{j+\delta_{2}',A\sigma}+t_{+}\bar{c}_{jB\sigma}c_{j+\delta_{2}',B\sigma})\nonumber\\
&+&\frac{J}{2}\sum_{j,\alpha}(\bar{c}_{j\alpha\uparrow}c_{j\alpha\downarrow}\bar{f}_{j\alpha\downarrow}f_{j\alpha\uparrow}+\bar{c}_{j\alpha\downarrow}c_{j\alpha\uparrow}\bar{f}_{j\alpha\uparrow}f_{j\alpha\downarrow})+\frac{J}{4}\sum_{j,\alpha,\sigma\sigma'}\sigma\sigma'\bar{c}_{j\alpha\sigma}c_{j\alpha\sigma}\bar{f}_{j\alpha\sigma'}f_{j\alpha\sigma'}-i\sum_{i,\alpha=A,B}\lambda_{i\alpha},
\end{eqnarray}
\end{widetext}
where $\lambda_{j\alpha}$ is a dynamic field whose effect is to decode the constraint $\hat{f}_{j\alpha\sigma}^{\dag}\hat{f}_{j\alpha\sigma}=1$. The partition function of the system is
\begin{equation}
\mathcal{Z}=\int \mathcal{D}\bar{c}\mathcal{D}c\mathcal{D}\bar{f}\mathcal{D}f\mathcal{D}\lambda e^{-S}.
\end{equation}
In order to match with mean-field theory in the main text, one uses the Hubbard-Stratonovich transformation and the resulting action reads,\cite{Coleman2015}
\begin{widetext}
\begin{eqnarray}
S&=&\int_{0}^{\beta}d\tau \sum_{j,\alpha=A,B,\sigma}\left(\bar{c}_{j\alpha\sigma}(\partial_{\tau}-\mu)c_{j\alpha\sigma}+\bar{f}_{j\alpha\sigma}(\partial_{\tau}+i\lambda_{j\alpha})f_{j\alpha\sigma}\right)-t\sum_{j,\delta,\sigma}(\bar{c}_{jA\sigma}c_{j+\delta,B\sigma}+\bar{c}_{j+\delta,B\sigma}c_{jA\sigma})\nonumber\\
&-&\sum_{j,\delta_{1}',\sigma}(t_{+}\bar{c}_{jA\sigma}c_{j+\delta_{1}',A\sigma}+t_{-}\bar{c}_{jB\sigma}c_{j+\delta_{1}',B\sigma})-\sum_{j,\delta_{2}',\sigma}(t_{-}\bar{c}_{jA\sigma}c_{j+\delta_{2}',A\sigma}+t_{+}\bar{c}_{jB\sigma}c_{j+\delta_{2}',B\sigma})\nonumber\\
&-&\frac{J}{2}\sum_{j,\alpha}(V_{j\alpha\uparrow}\bar{f}_{j\alpha\downarrow}c_{j\alpha\downarrow}+V_{j\alpha\downarrow}^{\ast}\bar{c}_{j\alpha\uparrow}f_{j\alpha\uparrow}-V_{j\alpha\downarrow}^{\ast}V_{j\alpha\uparrow})-\frac{J}{2}\sum_{j,\alpha}(V_{j\alpha\downarrow}\bar{f}_{j\alpha\uparrow}c_{j\alpha\uparrow}+V_{j\alpha\uparrow}^{\ast}\bar{c}_{j\alpha\downarrow}f_{j\alpha\downarrow}-V_{j\alpha\uparrow}^{\ast}V_{j\alpha\downarrow})\nonumber\\
&+&\frac{J}{4}\sum_{j,\alpha,\sigma}(\sigma m_{j\alpha}^{c}\bar{f}_{j\alpha\sigma}f_{j\alpha\sigma}+\sigma m_{j\alpha}^{f}\bar{c}_{j\alpha\sigma}c_{j\alpha\sigma}-m_{j\alpha}^{c}m_{j\alpha}^{f})-i\sum_{i,\alpha=A,B}\lambda_{i\alpha}.
\end{eqnarray}
\end{widetext}
Note that this action is invariant under $U(1)$ gauge transformation $f_{j\alpha\sigma}\rightarrow e^{i\theta_{j\alpha}}f_{j\alpha\sigma}$. Now, if all dynamic fields are setting to be static,
\begin{eqnarray}
&&V_{j\alpha\sigma}\rightarrow -V,~~i\lambda_{j\alpha}\rightarrow\lambda,~~m_{jA}^{c}\rightarrow m^{c},~~m_{jB}^{c}\rightarrow -m^{c}\nonumber\\
&&m_{jA}^{f}\rightarrow -m^{f},~~m_{jB}^{f}\rightarrow m^{f},
\end{eqnarray}
we recover the mean-field theory as expected.

When we discuss the phase transition between AM and the coexistent states, the magnetic order is intact but the Kondo order is critical, thus the corresponding action is
\begin{equation}
S=S_{0}+S_{V}\nonumber
\end{equation}
\begin{widetext}
\begin{eqnarray}
S_{0}&=&\int_{0}^{\beta}d\tau \sum_{k\sigma}\bar{c}_{kA\sigma}(\partial_{\tau}-\mu+\varepsilon_{k}^{AA}-Jm_{f}\sigma/2)c_{kA\sigma}+\sum_{k\sigma}\bar{c}_{kB\sigma}(\partial_{\tau}-\mu+\varepsilon_{k}^{BB}+Jm_{f}\sigma/2)c_{kB\sigma}\nonumber\\
&+&\sum_{k\sigma}\varepsilon_{k}(\bar{c}_{kA\sigma}c_{kB\sigma}+\bar{c}_{kB\sigma}c_{kA\sigma})+\sum_{k\sigma}\bar{f}_{kA\sigma}(\partial_{\tau}+\lambda+Jm_{c}\sigma/2)f_{kA\sigma}+\sum_{k\sigma}\bar{f}_{kB\sigma}(\partial_{\tau}+\lambda-Jm_{c}\sigma/2)f_{kB\sigma}
\end{eqnarray}
\end{widetext}
and
\begin{eqnarray}
S_{V}&=&\int_{0}^{\beta}d\tau\frac{J}{2}\sum_{j\alpha\sigma}\left(V_{j\alpha}\bar{f}_{j\alpha\sigma}c_{j\alpha\sigma}+V_{j\alpha}^{\ast}\bar{c}_{j\alpha\sigma}f_{j\alpha\sigma}+|V_{j\alpha}|^{2}\right)\nonumber\\
&=&\int_{0}^{\beta}d\tau\frac{J}{2}\sum_{kq\alpha\sigma}(V_{q\alpha}\bar{f}_{k+q,\alpha\sigma}c_{k\alpha\sigma}+V_{q\alpha}^{\ast}\bar{c}_{k-q,\alpha\sigma}f_{k\alpha\sigma}\nonumber\\
&+&\delta_{k=0}|V_{q\alpha}|^{2}).
\end{eqnarray}
Here, the spin index in the Kondo order field $V,V^{\ast}$ is neglected as guided by the mean-field theory. Generically, the interacting field theory described by $S=S_{0}+S_{V}$ cannot be solved exactly and further approximations has to be used.

In the sense of Landau symmetry-breaking theory, we may treat Kondo order field as the order parameter and its condensation leads to the coexistent state, where the Fermi surface is enlarged by hybridizing with Abriksov fermions.\cite{Senthil2004,Paul2007,Vojta2010,Zhong2012} In contrast, if $V,V^{\ast}$ do not condense, the system is in the AM state described by $S_{0}$ and the Abriksov fermions have no contribution to the Fermi surface.

When, Kondo order field is nearly critical, one can integrate out all fermionic degree of freedom and the resultant effective Gaussian action only including Kondo order field,
\begin{equation}
S_{eff}=\sum_{\Omega_{n},q}\left(J-(J/2)^{2}\chi_{0}^{\alpha}(q,\Omega_{n})\right)|V_{q\alpha}(\Omega_{n})|^{2}\label{eq:Ap4eq1}
\end{equation}
and the susceptibility $\chi_{0}^{\alpha}(q,\Omega_{n})$ is defined as
\begin{eqnarray}
\chi_{0}^{\alpha}(q,\Omega_{n})=-T\sum_{\omega_{n}}\sum_{k\sigma}G_{\alpha\sigma}^{c}(k,\omega_{n})G_{\alpha\sigma}^{f}(k+q,\omega_{n}+\Omega_{n}).\nonumber
\end{eqnarray}
The single-particle Green's functions for conduction electron $G_{\alpha\sigma}^{c}$ and Abriksov fermion $G_{\alpha\sigma}^{f}$ have the following expressions (with the help of $S_{0}$)
\begin{widetext}
\begin{eqnarray}
&&G_{A\sigma}^{c}(k,\omega_{n})=\frac{i\omega_{n}-(\varepsilon_{k}^{BB}-\mu+Jm_{f}\sigma/2)}{(i\omega_{n}-E_{k\sigma+})(i\omega_{n}-E_{k\sigma-})}=\frac{\mu_{k\sigma}^{2}}{i\omega_{n}-E_{k\sigma+}}+\frac{\nu_{k\sigma}^{2}}{i\omega_{n}-E_{k\sigma-}}\nonumber\\
&&G_{B\sigma}^{c}(k,\omega_{n})=\frac{\nu_{k\sigma}^{2}}{i\omega_{n}-E_{k\sigma+}}+\frac{\mu_{k\sigma}^{2}}{i\omega_{n}-E_{k\sigma-}}\nonumber\\
&&G_{A\sigma}^{f}(k,\omega_{n})=\frac{1}{i\omega_{n}-\lambda-Jm_{c}\sigma/2},~~G_{B\sigma}^{f}(k,\omega_{n})=\frac{1}{i\omega_{n}-\lambda+Jm_{c}\sigma/2}.
\end{eqnarray}
\end{widetext}
The coherent factors are defined as
\begin{equation}
\mu_{k\sigma}^{2}=1-\nu_{k\sigma}^{2}=\frac{E_{k\sigma+}-(\varepsilon_{k}^{BB}-\mu+Jm_{f}\sigma/2)}{E_{k\sigma+}-E_{k\sigma-}}.
\end{equation}
and the energy bands of quasiparticle in AM are
\begin{eqnarray}
E_{k\sigma\pm}&=&\frac{1}{2}\left(\varepsilon^{AA}_{k}+\varepsilon^{BB}_{k}\pm\sqrt{(\varepsilon_{k}^{AA}-\varepsilon_{k}^{BB}-Jm_{f}\sigma)^{2}+4\varepsilon^{2}_{k}}\right)\nonumber\\
&-&\mu.\nonumber
\end{eqnarray}
Inserting Green's function into the susceptibility $\chi_{0}^{\alpha}(q,\Omega_{n})$ gives the explicit expression of effective action of Kondo order field. If the Abriksov fermion has small dispersion (induced by Heisenberg interaction between local spins), one finds that the above action is similar to the one found in the study of Kondo-breakdown mechanism for frustrated $f$-electron materials.\cite{Senthil2004,Paul2007,Vojta2010,Zhong2012} However, it is still not clear that whether such Gaussian action is able to capture the true feature of this transition, where gapless fermions around Fermi surface couple to gapless bosons.\cite{Metlitski2010}

\section{Possibility of Striped order}\label{Ap5}
\begin{figure}
\includegraphics[width=1.2\linewidth]{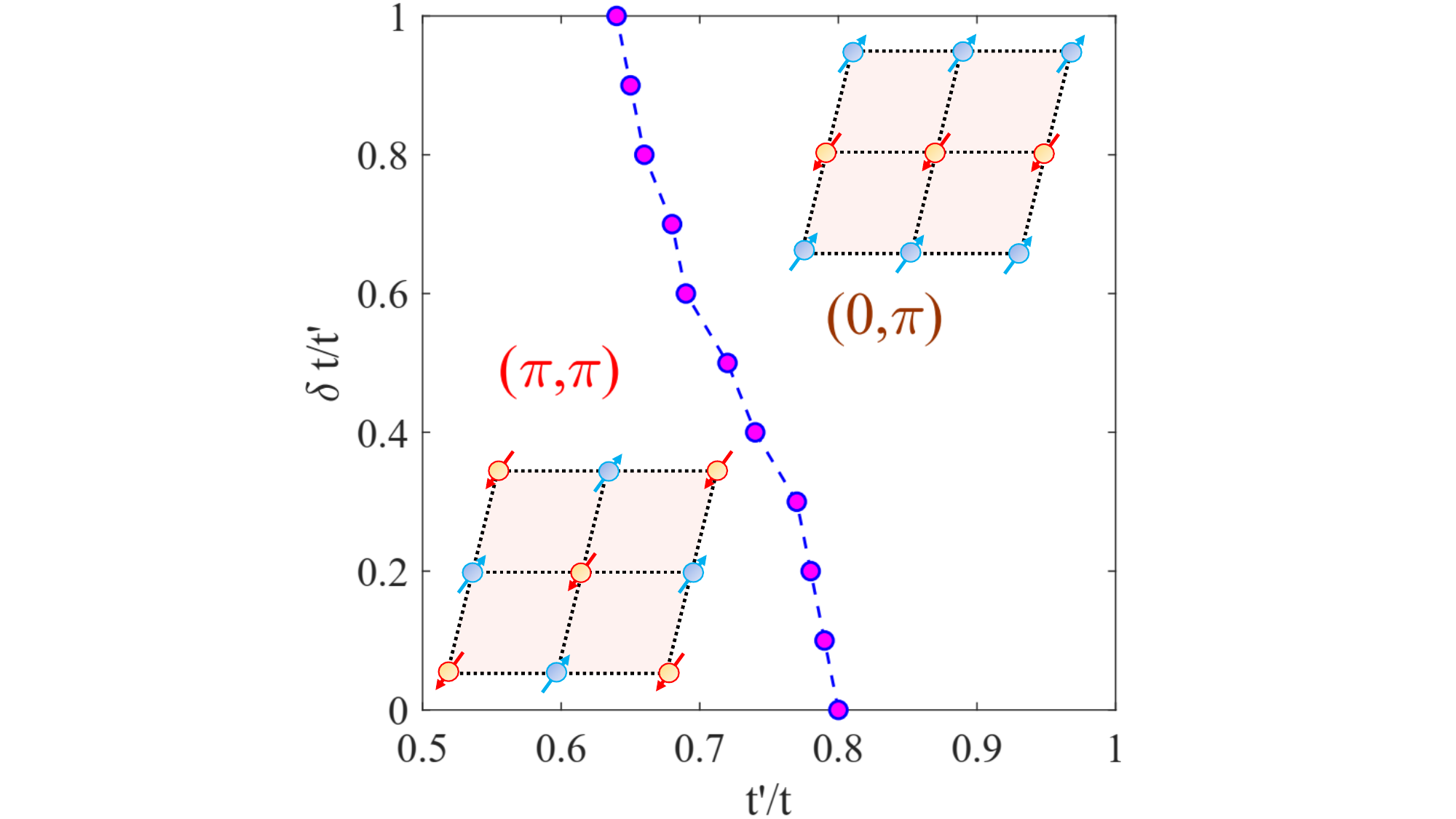}
\caption{\label{fig:13} The ground-state phase diagram at half-filling for $(\pi,\pi)$ order and $(0,\pi)$ order. ($t=1,J=1.2,n_{c}=1,T=0$)}
\end{figure}

Because there exists an finite NNNH in our Kondo lattice model, magnetic order beyond the antiferromagnetic Neel configuration ($(\pi,\pi)$-order) may appear. Here, we consider the striped order ($(\pi,0)$ or $(0,\pi)$ order), e.g. it is ferromagnetic in the $x$-direction but antiferromagnetic in the $y$-direction. In this case, the magnetic order parameters are chosen to be
\begin{eqnarray}
&&m_{A}^{c}(-1)^{j_{y}}=\frac{1}{2}\sum_{\sigma}\sigma\langle \hat{c}_{jA\sigma}^{\dag}\hat{c}_{jA\sigma}\rangle,m_{A}^{f}(-1)^{j_{y}}=\frac{1}{2}\sum_{\sigma}\sigma\langle \hat{f}_{jA\sigma}^{\dag}\hat{f}_{jA\sigma}\rangle,\nonumber\\
&&m_{B}^{c}(-1)^{j_{y}}=\frac{1}{2}\sum_{\sigma}\sigma\langle \hat{c}_{jB\sigma}^{\dag}\hat{c}_{jB\sigma}\rangle,m_{B}^{f}(-1)^{j_{y}}=\frac{1}{2}\sum_{\sigma}\sigma\langle \hat{f}_{jB\sigma}^{\dag}\hat{f}_{jB\sigma}\rangle.\nonumber
\end{eqnarray}
and
\begin{equation}
m_{A}^{c}=m_{B}^{c}=m_{c},~~m_{A}^{f}=m_{B}^{f}=-m_{f}.
\end{equation}
It is evident that the above magnetic order corresponds to the $(0,\pi)$-order. Inserting above formula into the mean-field decoupling of Kondo interaction, we find
\begin{eqnarray}
\hat{H}_{K}&=&\frac{J}{2}V\sum_{k\sigma}(\hat{c}_{kA\sigma}^{\dag}\hat{f}_{kA\sigma}+\hat{f}_{kA\sigma}^{\dag}\hat{c}_{kA\sigma})\nonumber\\
&+&\frac{J}{2}V\sum_{k\sigma}(\hat{c}_{kB\sigma}^{\dag}\hat{f}_{kB\sigma}+\hat{f}_{kB\sigma}^{\dag}\hat{c}_{kB\sigma})\nonumber\\
&+&\frac{J}{2}\sum_{k\sigma}\sigma(m_{c}\hat{f}^{\dag}_{kA\sigma}\hat{f}_{k+Q,A\sigma}-m_{f}\hat{c}_{kA\sigma}^{\dag}\hat{c}_{k+Q,A\sigma})\nonumber\\
&+&\frac{J}{2}\sum_{k\sigma}\sigma(m_{c}\hat{f}^{\dag}_{kB\sigma}\hat{f}_{k+Q,B\sigma}-m_{f}\hat{c}_{kB\sigma}^{\dag}\hat{c}_{k+Q,B\sigma})\nonumber\\
&+&2N_{s}J(V^{2}+m_{c}m_{f})\nonumber
\end{eqnarray}
with $Q=(0,\pi)$ as the characteristic wavevector of striped order. To proceed, $\hat{H}_{K}$ can be rewritten as
\begin{widetext}
\begin{eqnarray}
\hat{H}_{K}&=&\frac{J}{4}V\sum_{k\sigma}(\hat{c}_{kA\sigma}^{\dag}\hat{f}_{kA\sigma}+\hat{c}_{k+Q,A\sigma}^{\dag}\hat{f}_{k+Q,A\sigma}+h.c.)+\frac{J}{4}V\sum_{k\sigma}(\hat{c}_{kB\sigma}^{\dag}\hat{f}_{kB\sigma}+\hat{c}_{k+Q,B\sigma}^{\dag}\hat{f}_{k+Q,B\sigma}+h.c.)\nonumber\\
&+&\frac{J}{4}\sum_{k\sigma}\sigma(m_{c}\hat{f}^{\dag}_{kA\sigma}\hat{f}_{k+Q,A\sigma}-m_{f}\hat{c}_{kA\sigma}^{\dag}\hat{c}_{k+Q,A\sigma}+h.c.)+\frac{J}{4}\sum_{k\sigma}\sigma(m_{c}\hat{f}^{\dag}_{kB\sigma}\hat{f}_{k+Q,B\sigma}-m_{f}\hat{c}_{kB\sigma}^{\dag}\hat{c}_{k+Q,B\sigma}+h.c.)\nonumber\\
&+&2N_{s}J(V^{2}+m_{c}m_{f})\nonumber
\end{eqnarray}
\end{widetext}
At the same time, the non-interacting part has the following form,
\begin{widetext}
\begin{eqnarray}
\hat{H}_{0}&=&\frac{1}{2}\sum_{k\sigma}(\varepsilon_{k}\hat{c}_{kA\sigma}^{\dag}\hat{c}_{kB\sigma}+h.c.+\varepsilon_{k}^{AA}\hat{c}_{kA\sigma}^{\dag}\hat{c}_{kA\sigma}+\varepsilon_{k}^{BB}\hat{c}_{kB\sigma}^{\dag}\hat{c}_{kB\sigma})\nonumber\\
&+&\frac{1}{2}\sum_{k\sigma}(\varepsilon_{k+Q}\hat{c}_{k+Q,A\sigma}^{\dag}\hat{c}_{k+Q,B\sigma}+h.c.+\varepsilon_{k+Q}^{AA}\hat{c}_{k+Q,A\sigma}^{\dag}\hat{c}_{k+Q,A\sigma}+\varepsilon_{k+Q}^{BB}\hat{c}_{k+Q,B\sigma}^{\dag}\hat{c}_{k+Q,B\sigma}).\nonumber
\end{eqnarray}
\end{widetext}
Introducing spinor
\begin{widetext}
\begin{equation}
\hat{\psi}_{k\sigma}^{\dag}=(\hat{c}_{kA\sigma}^{\dag},\hat{c}_{kB\sigma}^{\dag},\hat{f}_{kA\sigma}^{\dag},\hat{f}_{kB\sigma}^{\dag},\hat{c}_{k+Q,A\sigma}^{\dag},\hat{c}_{k+Q,B\sigma}^{\dag},\hat{f}_{k+Q,A\sigma}^{\dag},\hat{f}_{k+Q,B\sigma}^{\dag}),
\end{equation}
\end{widetext}
the mean-field Hamiltonian reads,
\begin{equation}
\hat{H}=\sum_{k\sigma}\hat{\psi}_{k\sigma}^{\dag}H_{\sigma}(k)\hat{\psi}_{k\sigma}+2N_{s}(JV^{2}+Jm_{c}m_{f}-\lambda).
\end{equation}
where
\begin{equation}
H_{\sigma}(k)=\left(
                \begin{array}{cccc}
                  H^{cc}(k) & H^{cf} & H_{\sigma}^{ccQ} & 0 \\
                  H^{fc} & H^{ff} & 0 & H_{\sigma}^{ffQ} \\
                  H_{\sigma}^{ccQ} & 0 & H^{cc}(k+Q) & H^{cf} \\
                  0 & H_{\sigma}^{ffQ} & H^{fc} & H^{ff} \\
                \end{array}
              \right)
\end{equation}
\begin{widetext}
\begin{equation}
H^{cc}(k)=\frac{1}{2}\left(
                \begin{array}{cc}
                  \varepsilon_{k}^{AA}-\mu & \varepsilon_{k} \\
                  \varepsilon_{k} & \varepsilon_{k}^{BB}-\mu \\
                \end{array}
              \right),~~H^{ff}=\frac{1}{2}\left(
                \begin{array}{cc}
                  \lambda & 0 \\
                  0 & \lambda \\
                \end{array}
              \right),~~H^{cf}=H^{fc}=\frac{1}{2}\left(
                \begin{array}{cc}
                  \frac{JV}{2} & 0 \\
                  0 & \frac{JV}{2} \\
                \end{array}
              \right)
\end{equation}
\end{widetext}
and
\begin{eqnarray}
&&H_{\sigma}^{ccQ}=\frac{1}{2}\left(
                \begin{array}{cc}
                  -Jm_{f}\sigma/2 & 0 \\
                  0 & -Jm_{f}\sigma/2 \\
                \end{array}
              \right)\nonumber\\
&&H_{\sigma}^{ffQ}=\frac{1}{2}\left(
                \begin{array}{cc}
                 Jm_{c}\sigma/2 & 0 \\
                  0 & Jm_{c}\sigma/2 \\
                \end{array}
              \right).\nonumber
\end{eqnarray}
The order parameters $m_{c},m_{f},V$ are determined by
\begin{equation}
V=-\frac{1}{N_{s}}\sum_{k}\langle \hat{c}^{\dag}_{kA\sigma}\hat{f}_{kA\sigma}\rangle,
\end{equation}
\begin{equation}
m_{c}=\frac{1}{2N_{s}}\sum_{k\sigma}\sigma\langle \hat{c}^{\dag}_{kA\sigma}\hat{c}_{k+Q,A\sigma}\rangle,
\end{equation}
\begin{equation}
m_{f}=-\frac{1}{2N_{s}}\sum_{k\sigma}\sigma\langle \hat{f}^{\dag}_{kA\sigma}\hat{f}_{k+Q,A\sigma}\rangle.
\end{equation}
After obtaining $m_{c},m_{f},V$ by solving above equations, (equations for $n_{c}$ and $n_{f}$ are also needed) we compare the ground-state energy of the stripe order with the altermagnetic state. For example, in Fig.~\ref{fig:13}, the phase diagram in the half-filling case with $J/t=1.2$ has been established and we see that the altermagnetic state with $Q=(\pi,\pi)$ order is stable against the stripe order with $Q=(0,\pi)$ if the next-nearest-neighbor-hopping $t'$ is not larger than $0.6t$. ($t_{+}=t'+\delta t,t_{-}=t'-\delta t$)

\section{Green's function in coexistent state}\label{Ap6}
The mean-field Hamiltonian in the coexistent state is the following one
\begin{equation}
\hat{H}=\sum_{k\sigma}\hat{\psi}_{k\sigma}^{\dag}H_{\sigma}(k)\hat{\psi}_{k\sigma}+2N_{s}(JV^{2}+Jm_{c}m_{f}-\lambda)
\end{equation}
with the spinor $\hat{\psi}_{k\sigma}^{\dag}=(\hat{c}_{kA\sigma}^{\dag},\hat{c}_{kB\sigma}^{\dag},\hat{f}_{kA\sigma}^{\dag},\hat{f}_{kB\sigma}^{\dag})$ and $4\times4$-matrix
\begin{equation}
H_{\sigma}(k)=\left(
                \begin{array}{cc}
                  H_{\sigma}^{cc}(k) & H_{\sigma}^{cf}(k) \\
                  H_{\sigma}^{fc}(k) & H_{\sigma}^{ff}(k) \\
                \end{array}
              \right)
\end{equation}
\begin{equation}
H_{\sigma}^{cc}(k)=\left(
                \begin{array}{cc}
                  \varepsilon_{k}^{AA}-\mu-\frac{Jm_{f}\sigma}{2} & \varepsilon_{k} \\
                  \varepsilon_{k} & \varepsilon_{k}^{BB}-\mu+\frac{Jm_{f}\sigma}{2} \\
                \end{array}
              \right)
\end{equation}
\begin{equation}
H_{\sigma}^{ff}(k)=\left(
                \begin{array}{cc}
                  \lambda+\frac{Jm_{c}\sigma}{2} & 0 \\
                  0 & \lambda-\frac{Jm_{c}\sigma}{2} \\
                \end{array}
              \right)
\end{equation}
\begin{equation}
H_{\sigma}^{cf}(k)=H_{\sigma}^{fc}(k)=\left(
                \begin{array}{cc}
                  \frac{JV}{2} & 0 \\
                  0 & \frac{JV}{2} \\
                \end{array}
              \right)
\end{equation}
Therefore, the corresponding Green's function is
\begin{equation}
G_{\sigma}(k,\omega)=(\omega I_{4\times4}-H_{\sigma}(k))^{-1}
\end{equation}
which also has $4\times4$-matrix form
\begin{equation}
G_{\sigma}=\left(
             \begin{array}{cccc}
               G_{\sigma}^{cc,AA} & G_{\sigma}^{cc,AB} & G_{\sigma}^{cf,AA} & G_{\sigma}^{cf,AB} \\
               G_{\sigma}^{cc,BA} & G_{\sigma}^{cc,BB} & G_{\sigma}^{cf,BA} & G_{\sigma}^{cf,BB} \\
               G_{\sigma}^{fc,AA} & G_{\sigma}^{fc,AB} & G_{\sigma}^{ff,AA} & G_{\sigma}^{ff,AB} \\
               G_{\sigma}^{fc,BA} & G_{\sigma}^{fc,BB} & G_{\sigma}^{ff,BA} & G_{\sigma}^{ff,BB} \\
             \end{array}
           \right).
\end{equation}
In terms of above Green's function, we can calculate the Luttinger's integral
\begin{equation}
\mathrm{LI}_{\sigma}^{\alpha\beta,\Phi\Psi}=\frac{1}{N_{s}}\sum_{k}\theta(G_{\sigma}^{\alpha\beta,\Phi\Psi}(k,0)),
\end{equation}
which is for orbit index $\alpha,\beta=c,f$ and sublattice index $\Phi,\Psi=A,B$. Generally, if $\sum_{\sigma}\mathrm{LI}_{\sigma}^{\alpha\beta,\Phi\Psi}$ equals to the density of electron, the Luttinger theorem is conserved.
%

\end{document}